\documentclass[a4paper,titlepage,11pt]{article}
\pdfoutput=1
\usepackage[utf8]{inputenc}
\usepackage{geometry}
\usepackage[english]{babel}
\usepackage{amsmath, amsfonts, graphicx}
\usepackage{mathtools}
\usepackage{graphicx}
\usepackage{float}
\usepackage{mciteplus}
\usepackage{pstricks}
\usepackage{hyperref}
\usepackage[labelfont=bf,font={small}]{caption}
\usepackage{pgfplots}
\usepackage{empheq}
\usepackage{tikz}
\usetikzlibrary{positioning, arrows}
\usetikzlibrary{external}
\usepackage{footmisc}
\usepackage{multirow}
\usepackage{url}
\usepackage[breakable, theorems, skins]{tcolorbox}
\usepackage{enumerate}
\usepackage{physics}
\usepackage{bm}
\usepackage{epstopdf}

\newcommand*\widefbox[1]{\fbox{\hspace{0.5em}#1\hspace{0.5em}}}

\def\be{\begin{equation}}
\def\ee{\end{equation}}

\def\bea{\begin{eqnarray}}
\def\eea{\end{eqnarray}}

\newcommand{\beq}{\begin{equation}}
\newcommand{\eeq}{\end{equation}}
 
\newcommand{\barr}{\!\begin{array}}
\newcommand{\earr}{\end{array}\!}

\def\be{\bea}
\def\ee{\eea}

	\newcommand \mathtikz[1] {\quad \vcenter{\hbox{\tikz{#1}}} \quad}
	\newcommand\Int[2]{ 
\begin{scope}[xshift=#1,yshift=#2]
\draw (-0.35,.25) -- (0.35,.25);
\end{scope}}
	
	\numberwithin{equation}{section}

\begin{document}
\begin{titlepage}

\setcounter{page}{1} \baselineskip=15.5pt \thispagestyle{empty}

\vfil

${}$
\vspace{1cm}

\begin{center}

\def\thefootnote{\fnsymbol{footnote}}
\begin{changemargin}{0.05cm}{0.05cm} 
\begin{center}
{\Large \bf Quantum exponentials for the modular double and applications in gravity models}
\end{center} 
\end{changemargin}

~\\[1cm]
{Thomas G. Mertens\footnote{\href{mailto:thomas.mertens@ugent.be}{\protect\path{thomas.mertens@ugent.be}}}}
\\[0.3cm]

{\normalsize { \sl Department of Physics and Astronomy
\\[1.0mm]
Ghent University, Krijgslaan, 281-S9, 9000 Gent, Belgium}}\\[3mm]

\end{center}


 \vspace{0.2cm}
\begin{changemargin}{01cm}{1cm} 
{\small  \noindent 
\begin{center} 
\textbf{Abstract}
\end{center} 
In this note, we propose a decomposition of the quantum matrix group SL$_q^+(2,\mathbb{R})$ as (deformed) exponentiation of the quantum algebra generators of Faddeev's modular double of $\text{U}_q(\mathfrak{sl}(2, \mathbb{R}))$. The formula is checked by relating hyperbolic representation matrices with the Whittaker function. We interpret (or derive) it in terms of Hopf duality, and use it to explicitly construct the regular representation of the modular double, leading to the Casimir and its modular dual as the analogue of the Laplacian on the quantum group manifold. This description is important for both 2d Liouville gravity, and 3d pure gravity, since both are governed by this algebraic structure. This result builds towards a $q$-BF formulation of the amplitudes of both of these gravitational models.

}
\end{changemargin}
 \vspace{0.3cm}
\vfil
\begin{flushleft}
\today
\end{flushleft}

\end{titlepage}
\newpage
\tableofcontents

\setcounter{footnote}{0}

\section{Introduction}
Finding solvable models in lower dimensional gravity is a very important endeavor in our quest for understanding quantum gravity. In this light, the recent developments in Jackiw-Teitelboim (JT) gravity play an important role \cite{Jackiw:1984je, Teitelboim:1983ux, Almheiri:2014cka, Jensen:2016pah, Maldacena:2016upp, Engelsoy:2016xyb, Cotler:2016fpe, Stanford:2017thb, Kitaev:2018wpr, Mertens:2017mtv, Mertens:2018fds, Lam:2018pvp, Yang:2018gdb, Saad:2019lba,Mertens:2022irh}.

At the level of the action, JT gravity with negative cosmological constant can be reformulated in terms of a BF model based on the $\mathfrak{sl}(2,\mathbb{R})$ structure \cite{Fukuyama:1985gg, Isler:1989hq, Chamseddine:1989yz, Jackiw_1992}. One is hence led towards understanding amplitudes in JT gravity using the formalism of 2d BF theory, which is essentially an application of representation theory of the underlying algebraic structure. In particular, for the solution of amplitudes of 2d BF-models for group $G$, a crucial role is played by the group representation matrices $R_{mn}(g) = \langle R,m |\, g \, | R, n\rangle$. This is a direct consequence of the Peter-Weyl theorem implying the $R_{mn}(g)$ form an orthogonal basis for $L^2(G)$. These objects can physically be viewed as two-sided quantum wavefunctions $\hspace{-0.25cm} \mathtikz{\Int{0}{0};\node at (-.55,.25) {\small $m$};\node at (.55,.25)  {\small $n$};\node at (0,.45)  {\small $R$}}\hspace{-0.25cm}$ that diagonalize the quadratic Casimir, which for the models of interest is equal to the Hamiltonian of the system. This is in parallel to the solution strategy of 2d Yang-Mills models in the older literature \cite{Witten:1991we,Migdal:1975zg,Witten:1992xu}.

In some physical situations, the indices $m$ and/or $n$ at the boundaries are fixed and constrained to diagonalize one of the generators $X$ of the algebra $\mathfrak{g}$. The result is then a coset. E.g. if the ket is constrained as $X| R, n\rangle = 0$ for some generator $X\in\mathfrak{g}$, we obtain the right coset $G/U(1)$ where the 1-parameter subgroup generated by $X$ is modded out by right-multiplication. If both bra and ket are constrained as such, we have the double coset $U(1)\backslash G/U(1)$.

As mentioned, for lower-dimensional gravity with negative cosmological constant, the $\mathfrak{sl}(2,\mathbb{R})$ algebra is the relevant group-theoretical structure:
\begin{equation}
\label{eq:sl2r}
[H,E]=E,\quad [H,F]=-F,\quad [E,F] = 2H.
\end{equation}
For any representation, the representation matrix $R(g), \, g \in \text{SL}(2,\mathbb{R})$, can be written in the Gauss-Euler decomposition:
\begin{equation}
\label{eq:GE}
R(g) = e^{\gamma F} e^{2\phi H} e^{\beta E}, \qquad \gamma,\phi,\beta \in \mathbb{R}.
\end{equation}
Restricting the range of the coordinates $\gamma$ and $\beta$ to $\mathbb{R}^+$, $R(g)$ belongs to the positive subsemigroup SL$^+(2,\mathbb{R})$.

The asymptotic AdS boundary conditions \cite{Coussaert:1995zp,Bershadsky:1989mf} imply that the bra and ket of the relevant representation matrix elements are eigenstates of $F^\dagger$ and $E$ respectively (with eigenvalue $\nu$ and $\mu$ respectively that we will not further specify in this work).\footnote{We denoted these eigenstates with the label $\mathfrak{i}_L$ and $\mathfrak{i}_R$ in \cite{Blommaert:2018iqz} and subsequent work.} Hence the left- and rightmost element in the Gauss decomposition \eqref{eq:GE} are diagonalized on a two-boundary state and the important part of the matrix element reduces to only the Cartan element $e^{2\phi H}$ insertion as $R_{\nu\mu}(\phi) \equiv \langle \nu | e^{2\phi H} | \mu \rangle$ with principal series representation label $j=-1/2+ik,\, k \in \mathbb{R}^+$. This representation-theoretic object $R_{\nu\mu}(\phi)$ is called a \emph{Whittaker function} \cite{Jacquet,Schiffmann,Hashizume1,Hashizume2} or mixed parabolic matrix element, since its bra and ket diagonalize different parabolic generators. 
As the wavefunction on a slice that connects two asymptotic boundaries, it is the centerpiece of gravitational amplitudes in 2d JT gravity since it is used to compute the disk amplitude and boundary correlation functions \cite{Blommaert:2018oro,Blommaert:2018iqz}, see \cite{Fan:2021wsb} for the application to $\mathcal{N}=1$ JT supergravity, and \cite{Lin:2022zxd} for the $\mathcal{N}=2$ case.

However, JT gravity is not an isolated datapoint; it is instead connected to other exactly solvable gravitational models. We have seen in \cite{Fan:2021bwt,Mertens:2020hbs} that 2d Liouville gravity has a structure of the amplitudes that mirrors the JT case, but where the representation theoretic objects are $q$-deformed and doubled into the (tensor product) quantum algebra U$_q(\mathfrak{sl}(2,\mathbb{R})) \otimes $U$_{\tilde{q}}(\mathfrak{sl}(2,\mathbb{R}))$, where $q=e^{\pi i b^2}$ and $\tilde{q} = e^{\pi i b^{-2}}$.\footnote{This is a priori not surprising since there has been a long-standing research program identifying the modular tensor category of 2d Liouville/Virasoro CFT, to that of the modular double of U$_q(\mathfrak{sl}(2,\mathbb{R}))$ (see e.g. \cite{Teschner:2001rv}).} JT gravity then emerges in the classical $b\to 0$ scaling regime.\footnote{It should be mentioned that a different $q$-deformation of JT gravity also appears in the double-scaled SYK model \cite{Berkooz:2018qkz,Berkooz:2018jqr,Lin:2022rbf,Jafferis:2022wez,Goel:2023svz,Blommaert:2023opb}. The relevant $q$-deformation in that case is SU$_q(1,1)$ with $0<q<1$ a real number. In these works, the role of the two-sided wavefunctions is also emphasized.} 

Likewise, in \cite{Mertens:2022ujr}, see also \cite{Wong:2022eiu}, we presented arguments that 3d aAdS gravity in its interior is governed by representation theoretic objects consistent with a $q$-BF model based on the same structure U$_q(\mathfrak{sl}(2,\mathbb{R})) \otimes $U$_{\tilde{q}}(\mathfrak{sl}(2,\mathbb{R}))$, where the parameter $b$ is related to the AdS length $\ell$ through $c = 1+6(b+b^{-1})^2 \approx\frac{3\ell}{2G_N}$ in the semi-classical regime.\footnote{One actually has two such $q$-BF models: one for the holomorphic sector, and one for the antiholomorphic sector.} It was shown in the same work that taking $b\to 0$ combined with low temperature compared to the AdS length $\ell$, results in two copies of the JT description.

The combination of the quantum algebras with deformation parameter $q$ and $\tilde{q}$ is called the \emph{modular double}, since the parameter $b$ gets mapped to $1/b$, a modular $S$-transform.\footnote{Generalizations to the entire SL$(2,\mathbb{Z})$ family of deformed models $b\to \frac{Ab+B}{Cb+D}$ are possible, but these do not seem to be relevant for gravitational purposes.} The modular double of U$_q(\mathfrak{sl}(2,\mathbb{R}))$ was first introduced by L. Faddeev in 1999 \cite{Faddeev:1999fe} and received considerable attention since then. It was linked to 2d Liouville CFT and its fusion and braiding matrices in the seminal works \cite{Ponsot:1999uf,Ponsot:2000mt}, its defining properties were analyzed in more detail in \cite{Kharchev:2001rs,Bytsko:2002br,Bytsko:2006ut}, it was related to Teichm\"uller theory \cite{Nidaiev:2013bda,Teschner:2013tqy}, and has been extended by now to general split real quantum groups in a body of work \cite{Frenkel:2011nn,Ip,Ip2021}.

The importance of this quantum algebra cannot be overstated in lower-dimensional gravitational physics. The reason is that the quantum algebra that is related\footnote{By ``related to'', we mean the analogue of how for a compact Lie algebra $\mathfrak{g}$, the affine Lie algebra $\hat{\mathfrak{g}}$ is related to the quantum algebra $U_q(\mathfrak{g})$.} to the Virasoro algebra, is precisely the modular double U$_q(\mathfrak{sl}(2,\mathbb{R})) \otimes $U$_{\tilde{q}}(\mathfrak{sl}(2,\mathbb{R}))$. E.g. the Virasoro modular $S$-matrix 
\begin{equation}
S_0{}^{s} = 2\sqrt{2}\sinh(2\pi b s) \sinh (2\pi s/b),
\end{equation}
precisely matches with the Plancherel measure of the principal series representations of the modular double, which form a closed set of representations under tensor product. And in fact, one defines the modular double as the algebraic object that has precisely \emph{only} these irreducible representations \cite{Bytsko:2002br}. Since the Virasoro symmetry algebra is the governing principle for both 3d pure gravity, 2d Liouville gravity, and 2d JT gravity (as a limiting form of both of these), all of these gravitational models are governed in one way or another by this same underlying group-theoretic object.\footnote{See \cite{McGough:2013gka,Jackson:2014nla,Collier:2018exn} for some 3d gravitational applications that directly use the representation theory of the modular double.} \\

Our main goal in this note is to understand how the analogue of the Gauss-Euler structural decomposition of \eqref{eq:GE} works for this modular double algebraic structure relevant for 2d Liouville gravity and 3d gravity. Our proposal is the formula:
\begin{equation}
\label{eq:introfinal}
R(g) \, = \, g_{b}(\gamma f) \,  e^{2\phi H} \, g_b^*(\beta e),
\end{equation}
where $g_b$ is Faddeev's (non-compact) quantum dilogarithm \cite{Faddeev:1995nb}. The remaining notation will be explained in the main text. Finding such a formula is important since it is a step towards the ill-understood $q$-BF formulation (see e.g. \cite{Blau:1993tv, Aganagic:2004js,Szabo:2013vva}) of the amplitudes in both 2d Liouville gravity and 3d pure aAdS gravity.

This note is organized as follows. In \textbf{section \ref{s:constr}} we motivate and argue for the above formula \eqref{eq:introfinal} for the Gauss decomposition of the modular double. In \textbf{section \ref{s:hyperbolic}} we use this decomposition to explicitly compute the hyperbolic representation matrix element, finding agreement with an earlier determination by I. Ip, and providing evidence for our proposal. \textbf{Section \ref{s:appli}} describes some first applications in terms of the Casimir eigenvalue equation(s), the regular representation, and the interpretation of representation matrices as gravitational wavefunctions. \textbf{Appendix \ref{app:hopf}} provides a detailed technical description how the proposal \eqref{eq:introfinal} is a manifestation of Hopf duality applied directly to the modular double. A technical comment is made in \textbf{appendix \ref{app:A}}.

\section{Towards a Gauss-Euler decomposition of the modular double}
\label{s:constr}
If we write a generic $2\times 2$ quantum group element $g$ as 
\begin{equation}
g \equiv \left(\begin{array}{cc}A & B \\ C & D \end{array}\right) = \left(\begin{array}{cc} e^\phi & e^{\phi} \beta \\ \gamma e^\phi & \gamma e^\phi \beta + e^{-\phi} \end{array}\right),
\end{equation}
the quantum group SL$_q(2)$ is defined and parametrized by the variables $(A,B,C,D)$ satisfying:
\begin{gather} 
\label{sl2commutators}
AB= q BA, \qquad  AC= q CA,\qquad  BD= q DB,\qquad CD= q DC, \\
BC=CB,\qquad AD-DA = (q-q^{-1}) BC,
\end{gather}
or, equivalently, in terms of the non-commutative coordinates $(\gamma,\phi,\beta)$ satisfying
\begin{equation}
\label{eq:comm}
e^\phi\gamma = q \gamma e^\phi, \quad e^\phi \beta= q \beta e^\phi, \quad [\beta,\gamma]=0.
\end{equation}
For the case of the quantum group SL$_q(2)$, the Gauss-Euler decomposition in any representation is known, with a formula analogous as in \eqref{eq:GE}:\footnote{Note that the three generators separately do not define 1-parameter (quantum) subgroups, since setting two coordinates to zero is not always consistent with \eqref{eq:comm}. The case where only $\phi$ is non-zero is consistent, and that is the only exponential that is not $q$-deformed.}
\begin{equation}
\label{UqGE}
R(g) = e_{q^{-2}}^{\gamma F} \, e^{2\phi H} \, e_{q^2}^{\beta E}.
\end{equation}
This was first proven by C. Fronsdal and A. Galindo in \cite{Fronsdal:1991gf}, and further analyzed and extended in \cite{Bonechi:1993sn,Morozov:1994ab,Jagannathan:1994cm,VanDerJeugt:1995yn}. The generators $E,F$ and $H$ are representation matrices satisfying the dual U$_q(\mathfrak{sl}(2))$ quantum algebra:
\begin{equation}
\label{eq:alggm}
[H,E]=E,\quad [H,F]=-F,\quad [E,F] = \frac{q^{2H}-q^{-2H}}{q-q^{-1}}.
\end{equation}
For the two parabolic generators $E$ and $F$, the matrix exponential is $q$-deformed. The $q$-exponential is defined by its series expansion as:
\begin{equation}
e_q^x \equiv \sum_{n=0}^{+\infty} \frac{x^n}{[n]_q!}, \qquad [n]_q \equiv \frac{1-q^{n}}{1-q} = 1 + q + \hdots + q^{n-1},
\end{equation}
in terms of $q$-numbers $[n]_q$.
The equation \eqref{UqGE} is important since it relates the quantum algebra \eqref{eq:alggm} with the associated dual quantum group SL$_q(2)$, in a similar way as ordinary exponentiation relates a classical Lie algebra with a Lie group.

So far we have worked with the complex forms of these objects. Real forms of these quantum groups have been classified (see e.g. \cite{Klimyk:1997eb}) and consist of the compact real form SU$_q(2)$, the non-compact form SU$_q(1,1)$ where $q \in \mathbb{R}$, and the non-compact form SL$_q(2,\mathbb{R})$ with $\vert q \vert =1$. The compact real form SU$_q(2)$ is of no interest in gravity. The non-compact form SU$_q(1,1)$ is relevant for double-scaled SYK, but not for 3d gravity and Liouville gravity. We henceforth focus solely on the real form SL$_q(2,\mathbb{R})$. The reality condition in this case is enforced by the existence of a *-relation for which $A^*=A, \,B^*=B, \,C^*=C, \,D^*=D$. \\

Next, we work towards the modular double of SL$_q(2,\mathbb{R})$. The quantum algebra of the modular double U$_q(\mathfrak{sl}(2,\mathbb{R})) \otimes $U$_{\tilde{q}}(\mathfrak{sl}(2,\mathbb{R}))$ and its representations are constructed as follows. The irreducible representations of the modular double are continuous, and the generators can be represented as self-adjoint finite-difference operators acting on $L^2(\mathbb{R})$ as follows \cite{Ponsot:1999uf,Ponsot:2000mt,Hadasz:2013bwa}:
\begin{align}
K &\equiv q^H = T_{ib/2}, \nonumber \\
E &= e^{2\pi b t} \frac{e^{\pi ib \alpha}T_{ib/2} - e^{-\pi ib \alpha}T_{-ib/2}}{q-q^{-1}}, \label{PT} \\
F &= -e^{-2\pi b t} \frac{e^{-\pi ib \alpha}T_{ib/2} - e^{\pi ib \alpha}T_{-ib/2}}{q-q^{-1}}, \nonumber
\end{align}
where $T_a f(t) = f(t+a)$ is a shift operator. The representation label is $\alpha = Q/2 + i\mathfrak{s}$ for real $\mathfrak{s}$ and $Q=b+b^{-1}$. The algebra U$_q(\mathfrak{sl}(2,\mathbb{R}))$ is generated by the basis elements $F^lK^mE^n,\, l,n\in \mathbb{N},\, m \in \mathbb{Z}$.
One supplements to these three generators, the three dual generators $(\tilde{K}, \tilde{E}, \tilde{F})$ defined in terms of the same relations \eqref{PT} upon replacing $q \to \tilde{q}$ ($b\to b^{-1}$). The dual algebra is generated by the dual basis elements $\tilde{F}^l\tilde{K}^m\tilde{E}^n,\, l,n\in \mathbb{N}, \, m \in \mathbb{Z}$. The two pairs of generators $(K^2, E, F)$ and $(\tilde{K}^2, \tilde{E}, \tilde{F})$ commute with each other as can be readily checked. At the level of algebras, this means that the total algebra is a tensor product of both separate algebras, hence the notation U$_q(\mathfrak{sl}(2,\mathbb{R})) \otimes $U$_{\tilde{q}}(\mathfrak{sl}(2,\mathbb{R}))$. The resulting quantum algebra has by construction a symmetry $b\to b^{-1}$, unlike a single copy of U$_q(\mathfrak{sl}(2,\mathbb{R}))$, reflected in the Gauss decomposition of the dual quantum group \eqref{UqGE}. Finally, for the modular double it is natural to rescale the parabolic generators as
\begin{equation}
\label{eq:rescale}
e \equiv (2\sin\pi b^2) E, \qquad f \equiv (2\sin\pi b^2) F,
\end{equation}
since these parabolic generators have the crucial transcendental property \cite{Bytsko:2002br}: 
\begin{equation}
\label{eq:trans}
\tilde{e} \equiv (2\sin\pi b^{-2}) \tilde{E} = (e)^{1/b^2}, \qquad \tilde{f} \equiv (2\sin\pi b^{-2}) \tilde{F} = (f)^{1/b^2}.
\end{equation}

So far, the discussion was at the quantum algebra level. To pass to the modular double of the quantum group SL$_q(2,\mathbb{R})$, we will have to further adjust the $q$-exponentials in \eqref{UqGE} to implement the $b\to b^{-1}$ symmetry. In the spirit of Faddeev's original work \cite{Faddeev:1999fe}, it is clear that the correct replacement will be to use Faddeev's quantum dilogarithm $g_b$ (defined below) in place of the $q$-exponentials. We hence propose the following Gauss-Euler decomposition:
\begin{equation}
\label{eq:GEb}
\boxed{ R(g) \, = \, g_{b}(\gamma f) \,  e^{2\phi H} \, g_b^*(\beta e)},
\end{equation}
as the bridge between the quantum algebra U$_q(\mathfrak{sl}(2,\mathbb{R})) \otimes $U$_{\tilde{q}}(\mathfrak{sl}(2,\mathbb{R}))$ and the modular double quantum group SL$_q^+(2,\mathbb{R})$. The triple $(\gamma,\phi,\beta)$ satisfy the same commutator relations \eqref{eq:comm} as before, and where $\gamma$ and $\beta$ are naturally restricted to a positive spectrum, since they appear as arguments within $g_b$ which incorporates non-polynomial powers of its argument in its definition. This explains the superscript ${}^+$ for the associated quantum group SL$_q^+(2,\mathbb{R})$.

Let us motivate the proposal \eqref{eq:GEb} in more detail. The relevant deformed exponential is Faddeev's quantum dilogarithm $g_b(x)$, defined as \cite{Faddeev:1995nb,Faddeev:1994fw}:
\begin{equation}
g_b(x) \equiv \exp\left(- \int_{\mathbb{R}+i0} \frac{dt}{4t} \frac{x^{\frac{t}{2\pi i b}}}{\sinh \frac{bt}{2} \sinh \frac{t}{2b}}\right).
\end{equation}
It satisfies the properties:
\begin{align}
g_b(u+v) &= g_b(u)g_b(v), \qquad uv = q^2 vu, \\
\vert g_b(x) \vert &=1 \text{ if $x \in \mathbb{R}^+$}, \\
g_b(x^b) &= g_{b^{-1}}(x^{b^{-1}}).
\end{align}
The first two properties identify $g_b(x)$ as a quantum exponential function. The last one implements the self-duality $(b\to b^{-1})$ at the level of the quantum group representation matrices in \eqref{eq:GEb}. The function $g_b$ has the following inverse Mellin transform in terms of the double sine function $S_b(x)$:\footnote{For the definition of the double sine function $S_b(x)$, see e.g. appendix B of \cite{Fan:2021bwt} where we compiled some properties.}
\begin{align}
\label{eq:eb}
g_b(x)  &= -ib\int_{i\mathbb{R}} ds (-ix)^{-s} S_b(bs)q^{\frac{s^2}{2}+\frac{s}{2}}, \\
\label{eq:eb2}
g_b^*(x) &=-ib\int_{i\mathbb{R}} ds (ix)^{-s} S_b(bs)q^{-\frac{s^2}{2}-\frac{s}{2}}.
\end{align}
We will evaluate these integrals by contour deforming to the left half-plane where we pick up the residues of all poles of the $S_b$-function. The $S_b$-function has known double sets of poles with residues:
\begin{align}
\text{Res}\left.S_b\right|_{x=-mb-n/b} &= \frac{S_b(Q)}{2\pi}\frac{(-)^{n+m+nm}}{S_b(mb+n/b+b+1/b)} = \frac{1}{2\pi}\frac{(-)^{n+m+nm}}{\prod_{i=1}^{m}2 \sin \pi b^2 i \prod_{j=1}^{n}2 \sin \pi b^{-2} j} \nonumber \\
&= \frac{1}{2\pi}(-)^{nm} (2\sin \pi b^2)^{-m}(2\sin \pi b^{-2})^{-n}\frac{(-)^m}{[\![m]\!]_q!} \frac{(-)^n}{[\![n]\!]_{\tilde{q}}!},
\end{align}
where $(n,m) \in \mathbb{N}^2$ (including zero), and where in the last line we introduced the ``symmetric'' $q$-number as:
\begin{equation}
[\![n]\!]_q = \frac{q^n-q^{-n}}{q-q^{-1}}.
\end{equation}
So the $g_b$-function has the double series expansion:
\begin{align}
g_b(x) &= \sum_{n,m} \frac{i^m}{q^{-\frac{m(m-1)}{2}}[\![m]\!]_q!} \frac{i^n}{\tilde{q}^{-\frac{n(n-1)}{2}}[\![n]\!]_{\tilde{q}}!} (2\sin \pi b^2)^{-m}(2\sin \pi b^{-2})^{-n} x^{m+n/b^2}.
\end{align}
Using the identity:
\begin{equation}
e_{q^2}^x = \sum_{n=0}^{+\infty} \frac{x^n}{[n]_{q^2}!} = \sum_{n=0}^{+\infty} \frac{q^n x^n}{q^{\frac{n(n+1)}{2}}[\![n]\!]_q!},
\end{equation}
we can rewrite this suggestively as a product of two $q$-exponentials as:
\begin{align}
g_b(x) = e_{q^{-2}}^{\frac{ix}{2\sin \pi b^2}} e_{\tilde{q}^{-2}}^{\frac{i x^{1/b^2}}{2\sin \pi b^{-2}}}.
\end{align}
Using the rescaled generators \eqref{eq:rescale} and the transcendental relations \eqref{eq:trans}, we have:
\begin{align}
\label{eq:bexp}
g_b(\gamma f) &= e_{q^{-2}}^{i \gamma F} \, e_{\tilde{q}^{-2}}^{i \gamma^{1/b^2}\tilde{F}}, \\
g_b^*(\beta e) &= e_{q^{2}}^{-i \beta E} \, e_{\tilde{q}^2}^{-i \beta^{1/b^2}\tilde{E}},
\end{align}
containing simultaneously both a generator from the quantum group and its dual, in the same exponentiated form as in \eqref{UqGE}. Plugging this in \eqref{eq:GEb}, we can expand
\begin{equation}
\label{eq:expge}
R(g) \, = \, e_{\tilde{q}^{-2}}^{i \gamma^{1/b^2}\tilde{F}} \, e_{q^{-2}}^{i \gamma F} \,  e^{2\phi H} \, e_{q^{2}}^{-i \beta E} \, e_{\tilde{q}^2}^{-i \beta^{1/b^2}\tilde{E}},
\end{equation}
which contains a product of five exponentials. 

Comparing to \eqref{UqGE}, we note that there are factors of $i$ inserted in the deformed exponentials. This is no surprise; we have explained in earlier work \cite{Blommaert:2018iqz,Fan:2021wsb} that the eigenvalues of the $E$ and $F$ generators acquire additional factors of $\pm i$ when transferring from the full group SL$(2,\mathbb{R})$ to the positive semigroup SL$^+(2,\mathbb{R})$, which is a reflection of this property on the undeformed $q\to 1$ limit.

Notice the Cartan generator $H$ is \emph{not} doubled in \eqref{eq:expge}. This is because, after setting $\phi \to b \phi$ such that $e^{2\phi b H} = e^{2\phi b^{-1} \tilde{H}}$, it is $b\to b^{-1}$ invariant and hence ``serves both quantum groups'' in Faddeev's wording \cite{Faddeev:1999fe}. A perhaps more natural choice of parametrization of the modular double is to hence rescale $\phi \to b \phi$, $\gamma,\beta \to \gamma^b, \beta^b$, leading to the representation matrix:
\begin{equation}
\label{eq:rescrep}
R(g) \, = \, e_{\tilde{q}^{-2}}^{i \gamma^{1/b}\tilde{F}} \, e_{q^{-2}}^{i \gamma^b F} \,  e^{\phi b H} \, e_{q^{2}}^{-i \beta^b E} \, e_{\tilde{q}^2}^{-i \beta^{1/b}\tilde{E}},
\end{equation}
where one has the tantalizingly simple non-commutativity relations:
\begin{equation}
e^\phi\gamma = -\gamma e^\phi, \quad e^\phi \beta= - \beta e^\phi, \quad [\beta,\gamma]=0.
\end{equation}

We develop the interpretation of the formula \eqref{eq:expge} in terms of duality of Hopf algebras in Appendix \ref{app:hopf}, and argue in more detail why a representation of the modular double quantum algebra maps into a representation of the associated dual matrix quantum group. This is what ordinarily constitutes a proof of the above formula \eqref{eq:expge}, see \cite{Fronsdal:1991gf}. However, one can read our arguments there also as an interpretation on what the modular double quantum group is concretely as a Hopf algebra.

In the next section, we apply \eqref{eq:GEb} directly to compute some important representation matrix elements, matching them with earlier results and hence providing indirect evidence for the validity of \eqref{eq:GEb}.

\section{$q$-Representation matrix elements}
As mentioned in the Introduction, representation matrix elements $R_{mn}(g)$ can be written in a bra-ket notation as $\left\langle m \right| g \left|n \right\rangle$. Different bases for the bra or ket indices $m$ and $n$ respectively can then be related by inserting complete sets of states as
\begin{equation}
\label{insbasis}
R_{\nu\mu}(g) \equiv \left\langle \nu \right| g \left|\mu\right\rangle  = \int_{-\infty}^{+\infty} ds_1 ds_2 \left\langle  \nu \right| \left. s_1\right\rangle \left\langle s_1 \right| g \left|s_2\right\rangle \left\langle s_2 \right| \left. \mu\right\rangle.
\end{equation}
We will use the greek $\nu$ and $\mu$ indices to represent so-called parabolic indices, diagonalizing the parabolic generators $F^\dagger$ and $E$ of \eqref{eq:sl2r} respectively. The latin $s_{1,2}$ indices denote hyperbolic indices diagonalizing the hyperbolic generator $K=q^H$ instead. The above change-of-basis relation is well-known for classical Lie groups \cite{VK}, but here we explore it for the modular double quantum group at hand.

\label{s:hyperbolic}
\subsection{Warm-up: Whittaker function}
As warm-up for the generic representation matrix element of the next section, we compute here the Whittaker function for the modular double, where $g$ is in the Cartan subgroup. This result has been known for some time due to work by S. Kharchev, D. Lebedev and M. Semenov-Tian-Shansky \cite{Kharchev:2001rs}, but we perform the computation in a slightly different way utilizing \eqref{insbasis}, which will allow us to leverage it towards computing the generic hyperbolic matrix element in the next subsection. In gravity, the Whittaker function describes the two-sided wavefunction as described in the Introduction.

In the current set-up, $g= e^{2\phi H}$. The hyperbolic generator $K = T_{ib/2} = q^H$ where $H = \frac{1}{2\pi b}\partial_t$, can be diagonalized as 
\begin{equation}
K \psi(t) = e^{-\pi b^2 s} \psi(t), \qquad \tilde{K} \equiv K^{1/b^2} =  e^{-\pi s} \psi(t), \qquad s \in \mathbb{R},
\end{equation}
with solution 
\begin{equation}
\label{eq:Hwv}
\phi_s(t) \equiv \left\langle t \right| \left. s \right\rangle = \sqrt{b}\, e^{2\pi i b s t},\,\, \text{   or   } \,\, \left\langle x \right| \left. s \right\rangle\frac{1}{\sqrt{2\pi}} x^{is},
\end{equation}
and eigenvalue $H \psi(t) = is \psi(t)$, where we have set $x=e^{2\pi b t}$. These modes are orthonormal:
\begin{equation}
\int_{-\infty}^{+\infty} dt \psi^*_{s_1}(t)\psi_{s_2}(t) = \delta(s_1-s_2).
\end{equation}
The resulting hyperbolic matrix element with only the Cartan generator inserted is given by
\begin{equation}
\label{cartel}
\left\langle s_1 \right| e^{2 \phi H} \left| s_2\right\rangle = e^{2i s_1 \phi} \delta(s_1-s_2).
\end{equation}
The Whittaker vector $\phi_\mu(t)$ simultaneously diagonalizing $E$ and $\tilde{E}$,
\begin{equation}
\label{eq:whitv1}
E \phi_\mu(t) = \frac{i \mu}{q-q^{-1}} \phi_\mu(t), \qquad \tilde{E} \phi_\mu(t) = \frac{i \mu^{1/b^2}}{\tilde{q}-\tilde{q}^{-1}} \phi_\mu(t), \quad \mu \geq 0,
\end{equation}
was determined in \cite{Kharchev:2001rs}.\footnote{A more general eigenvalue problem was considered there with the action of the Cartan generator on the RHS as $q^{2\alpha_2 H}$, for $\alpha_2$ an additional parameter allowed in $q$-deforming the eigenvalue problem. We set this $\alpha_2=0$ here since we will not need it to make our point. However, for the gravitational application to Liouville gravity, this parameter is actually non-zero \cite{Fan:2021bwt,Mertens:2020hbs} and hence a slightly more general matrix element has to be considered. We will turn on this parameter in subsection \ref{sec:onesided}.} Note that this system is consistent, with these precise prefactors, due to the transcendental relation \eqref{eq:trans}, implying we are in fact diagonalizating $e$ and $\tilde{e}\equiv e^{1/b^2}$. Using the parametrization \eqref{PT} of the quantum algebra, the Whittaker vector is\footnote{In \cite{Fan:2021bwt}, we determined it with the replacement $\alpha = Q/2 + i\mathfrak{s} \to \lambda =b/2 + i\mathfrak{s}$, and using anti-self-adjoint operators instead, again requiring a slightly different parametrization. That parametrization was convenient to analyze the classical $b\to 0$ limit. This is not our focus in this work.}
\begin{equation}
\label{eq:whE}
\phi_\mu(t) \equiv \left\langle t\right|\left. \mu \right\rangle= e^{-2\pi \alpha t} \int_\mathcal{C} d\zeta \mu^{i\zeta/b} S_b(-i\zeta) e^{-2\pi i \zeta t},
\end{equation}
where $\alpha = Q/2 + i\mathfrak{s}$ and where the integration contour $\mathcal{C}$ is along the real axis above the pole at the origin $\zeta=0$. \\

From \eqref{eq:Hwv} and \eqref{eq:whE} we can compute the overlap between these states as
\begin{align}
\label{overl1}
\left\langle s_2\right|\left. \mu \right\rangle \equiv 2\pi b \int dt \left\langle s_2\right|\left. t \right\rangle \left\langle  t \right| \left. \mu \right\rangle = b^{1/2} \, \mu^{-is_2 -\alpha/b} S_b(ibs_2 + \alpha).
\end{align}

It is useful to compare this overlap to its classical limit. To find it, we rescale 
\begin{align}
\phi_\mu(t) \to \sqrt{2\pi b} e^{+\pi \frac{t}{b}} \phi_\mu(t), \qquad \mu \to 2\pi b^2 \mu, \qquad \alpha = Q/2-ibk,
\end{align} 
where the new $\mu$ and $k$ are fixed as $b\to 0$. Using $S_b(bx) \underset{b\to 0}{\to} \frac{1}{\sqrt{2\pi}} (2\pi b^2)^{x-1/2}\Gamma(x)$, we find the classical $b\to 0$ overlap:
\begin{align}
\left\langle s_2\right|\left. \mu \right\rangle \to \frac{1}{\sqrt{2\pi}} \Gamma(1/2 + is_2-ik) \mu^{-is_2+ik-1/2}.
\end{align}

Likewise, for the conjugate of the Whittaker vector simultaneously diagonalizing $F^\dagger$ and $\tilde{F}^\dagger$:\footnote{Also here, a further parameter $\alpha_1$ is set to zero in the eigenvalue problem in a similar manner. For the application to Liouville gravity, one in the end wants to set $\epsilon\equiv \alpha_1-\alpha_2 = \pm 1$, which due to symmetry requires $\alpha_1 = - \alpha_2 = \pm 1/2$. It would be interesting to explain this choice of boundary condition directly from the gravitational variables in Liouville gravity.}
\begin{equation}
\label{eq:whitv2}
F^\dagger \phi_\nu(t) = \frac{i \nu}{q-q^{-1}} \phi_\nu(t), \qquad \tilde{F}^\dagger \phi_\mu(t) = \frac{i \nu^{1/b^2}}{\tilde{q}-\tilde{q}^{-1}} \phi_\nu(t) , \quad \nu \geq 0,
\end{equation}
the expression:
\begin{equation}
\bar{\phi}_\nu(t) \equiv \left\langle \nu\right|\left. t \right\rangle= e^{2\pi \bar{\alpha} t} \int_\mathcal{C} d\zeta \nu^{-i\zeta/b} S_b(i\zeta) e^{-2\pi i \zeta t},
\end{equation}
leading to the overlap:
\begin{align}
\label{overl2}
\left\langle \nu \right|\left. s_1 \right\rangle &\equiv 2\pi b \int dt \left\langle \nu \right|\left. t \right\rangle \left\langle  t \right| \left. s_1 \right\rangle = b^{1/2} \, \nu^{-is_1 -\bar{\alpha}/b} S_b(ibs_1 + \bar{\alpha}),
\end{align}
with analogous classical $b\to 0$ limit (after suitable rescalings again):
\begin{align}
\left\langle \nu \right|\left. s_1 \right\rangle \to \frac{1}{\sqrt{2\pi}} \Gamma(1/2 + is_1 + ik) \nu^{-is_1-ik-1/2}.
\end{align}

Inserting the different ingredients \eqref{cartel}, \eqref{overl1} and \eqref{overl2}, we obtain:
\begin{align}
\label{eq:whit}
R_{\nu\mu}(\phi) \equiv \left\langle \nu \right| e^{2\phi H} \left|\mu\right\rangle &= e^{-Q \frac{\phi}{b}} e^{-2 i \mathfrak{s} \frac{\phi}{b}} \int_{\mathcal{C}} d\zeta\, \mu^{i\zeta/b} \nu^{i\zeta/b + 2is/b}S_b(-i\zeta) S_b(-i\zeta-2is)e^{-2 i \zeta \frac{\phi}{b}}  \nonumber \\
&= \int_{\mathcal{C}} d\zeta\, \mu^{-i\frac{\zeta}{b}-\frac{\alpha}{b}} \nu^{-i\frac{\zeta}{b} - \frac{\bar{\alpha}}{b}}S_b(i\zeta+\alpha) S_b(i\zeta+\bar{\alpha})e^{2 i \zeta \frac{\phi}{b}},
\end{align}
indeed matching with the Whittaker function of \cite{Kharchev:2001rs} in the specific case where $\epsilon=0$.\footnote{The prefactor $e^{- \pi Q x}$ in the first line is sometimes explicitly removed, and absorbed in the Haar measure on the quantum group manifold. We do not perform this step here.} 

\subsection{Hyperbolic representation matrix element}
We start anew with the equality
\begin{equation}
\label{eq:main}
R_{\nu\mu}(g) = \left\langle \nu \right| g \left|\mu\right\rangle  = \int_{-\infty}^{+\infty} ds_1 ds_2 \left\langle \nu \right| \left. s_1\right\rangle \left\langle s_1 \right| g \left|s_2\right\rangle \left\langle s_2 \right| \left. \mu\right\rangle,
\end{equation}
but now insert the full quantum group element $g$ in the form of our proposal \eqref{eq:GEb}. 

On the LHS of \eqref{eq:main}, the ket and bra diagonalize the rescaled self-adjoint parabolic generators $e$ and $f$ with eigenvalue $\nu$ and $\mu$ respectively.
So the LHS can be simplified into
\begin{align}
\label{eq:lhs}
&g_{b}(\gamma\nu) \left\langle R, \nu \right| e^{2\phi H} \left|R,\mu\right\rangle  g_b^*(\beta \mu)\nonumber \\
&= g_{b}(\gamma\nu)\int_{\mathcal{C}} d\zeta\, \mu^{-i\frac{\zeta}{b}-\frac{\alpha}{b}} \nu^{-i\frac{\zeta}{b} - \frac{\bar{\alpha}}{b}}S_b(i\zeta+\alpha) S_b(i\zeta+\bar{\alpha})e^{2\pi i \zeta x} g_b^*(\beta \mu),
\end{align}
where we inserted the Whittaker function \eqref{eq:whit}. 

To evaluate the RHS of \eqref{eq:main}, we can use the expressions \eqref{overl1} and \eqref{overl2}:
\begin{align}
\left\langle s_2\right|\left. \mu \right\rangle &= b^{1/2} \, \mu^{-is_2 -\alpha/b} S_b(ibs_2 + \alpha), \\
\left\langle \nu \right|\left. s_1 \right\rangle &= b^{1/2}  \, \nu^{-is_1 -\bar{\alpha}/b} S_b(ibs_1 + \bar{\alpha}).
\end{align}
Inserting these in \eqref{eq:main}, and performing the integral transforms (ranging only over positive values of $\mu$ and $\nu$):
\begin{align}
\label{eq:inttrans}
\int_0^{+\infty} d\mu \mu^{i\tilde{s}_2-1/2+1/2b^2+i\mathfrak{s}/b} \int_0^{+\infty} d\nu \nu^{i\tilde{s}_1-1/2+1/2b^2-i\mathfrak{s}/b} \times \hdots
\end{align}
on both sides, the RHS reduces to
\begin{equation}
(2\pi)^2 b \, S_b(ib\tilde{s}_1 + \bar{\alpha}) S_b(ib\tilde{s}_2 + \alpha) \left\langle R, \tilde{s}_1 \right| g \left|R, \tilde{s}_2\right\rangle.
\end{equation}
Hence we obtain the hyperbolic representation matrix element:
\begin{align}
\label{eq:prefinal}
&\left\langle \tilde{s}_1 \right| g \left|\tilde{s}_2\right\rangle \equiv R_{\tilde{s}_1\tilde{s}_2}(g) = \frac{1}{(2\pi)^2 bS_b(ib\tilde{s}_2 + \alpha) S_b(ib\tilde{s}_1 + \bar{\alpha})} \, \\
&\times \int_{\mathcal{C}} d\zeta \int_0^{+\infty} \frac{d\nu}{\nu} \nu^{i(\tilde{s}_1-\zeta/b)} g_{b}(\gamma\nu) S_b(i\zeta+\alpha) S_b(i\zeta+\bar{\alpha}) e^{2 i \zeta \frac{\phi}{b}} \int_0^{+\infty} \frac{d\mu}{\mu} \mu^{i(s_2-\zeta/b)} g_b^*(\beta \mu). \nonumber
\end{align}
Evaluating the integrals using the Mellin transform of \eqref{eq:eb}, we finally obtain (after dropping the tildes):
\begin{empheq}[box=\widefbox]{align*}
&R_{s_1s_2}(g) =N \nonumber \\
&\times \gamma^{-is_1} \hspace{-0.2cm}\int_{\mathcal{C}} d\zeta\, \frac{S_b(ibs_2-i\zeta) S_b(ibs_1-i\zeta)S_b(i\zeta+\alpha) S_b(i\zeta+\bar{\alpha})}{S_b(ibs_2 + \alpha) S_b(ibs_1 + \bar{\alpha})} (\gamma e^{2 \phi}\beta)^{i \zeta /b} e^{\pi i b (s_1-s_2)\zeta} \beta^{-is_2}
\end{empheq}
\begin{equation}\label{eq:final}
\end{equation}
with the explicit normalization factor
\begin{equation}
N = b e^{\frac{\pi}{2}(s_2-s_1)}e^{\frac{\pi i b^2}{2}(s_2^2-s_1^2)}e^{\frac{\pi b^2}{2}(s_2-s_1)}.
\end{equation}
We used the following equality:
\begin{equation}
\gamma^z e^{2z\phi} \beta^z = (\gamma e^{2\phi}\beta)^{z}, \quad z \in \mathbb{C}.
\end{equation}
The quantity $\gamma e^{2\phi}\beta$ is sometimes called the hyperbolic element \cite{VK}, and is the only combination of the coordinates the integral in \eqref{eq:final} depends on.

One can see here that if we picked the ``wrong'' deformed exponentials from \eqref{UqGE} to compute the representation matrix element, we would produce instead $q$-Gamma functions as a result of the $\mu$- and $\nu$-integrals (see Appendix \ref{app:A}).

In gravity, the hyperbolic representation matrix element is relevant for describing fully internal wavefunctions (whose endpoints are \emph{not} on holographic boundaries).

\subsection{Comparison to earlier work}

The correctness of our result can be appreciated by comparing \eqref{eq:final} to an earlier determination of this hyperbolic representation matrix element using a different technique to which we turn next. In particular, our result \eqref{eq:final} is to be compared to the expression (7.35) of I. Ip \cite{Ip}. We first rewrite our expression slightly. 
Using
\begin{equation}
(e^\phi)^{2i\zeta /b}\beta^{-is_2} = e^{2\pi i b s_2\zeta} \beta^{-is_2} (e^\phi)^{2i\zeta /b},
\end{equation}
the substitution $i\zeta = -i\tau - \alpha$ and the identity $S_b(x) = 1/S_b(Q-x)$, we can rewrite \eqref{eq:final} as
\begin{align}
\label{eq:toco}
&\tilde{N} \gamma^{-is_1} \beta^{-is_2} (\gamma e^{2 \phi}\beta)^{-\alpha /b} \\
&\hspace{-0.15cm}\times\frac{S_b(\alpha-ibs_1)}{S_b(\alpha+ ibs_2)}\int_{\mathcal{C}} d\tau\, \frac{S_b(\alpha + ibs_2+i\tau) S_b(\alpha + ibs_1+i\tau)S_b(-i\tau)}{S_b(2\alpha + i\tau)} (\gamma e^{2 \phi}\beta)^{-i \frac{\tau}{b}} e^{-\pi i  (bs_1+bs_2)\tau}, \nonumber
\end{align}
with a new normalization factor $\tilde{N}$ that we will not track explicitly. 

Let us now compare this expression to that of \cite{Ip}. Within the ``quantum plane'' decomposition of a GL$_q^+(2,\mathbb{R})$ matrix as:
\begin{equation}
\left(\begin{array}{cc}z_{11} & z_{12} \\ z_{21} & z_{22} \end{array}\right) = \left(\begin{array}{cc}A & 0 \\ B & 1 \end{array}\right) \left(\begin{array}{cc}1 & \hat{B} \\ 0 & \hat{A} \end{array}\right),
\end{equation}
where $AB = q^2 BA$ and $\hat{A}\hat{B} = q^{-2}\hat{B}\hat{A}$ and $\text{qdet} = z_{11}z_{22}-z_{12}z_{21} = A\hat{A}$, the hyperbolic representation matrix element was determined using a different technique in \cite{Ip}:\footnote{We set $s_{\text{there}} \to -bs_1$, $\alpha_{\text{there}} = -bs_2$ and $l_\text{there} = -\alpha$.}
\begin{align}
R_{s_1s_2}^\alpha(g) &= A^{-\frac{\alpha}{b}+is_1}B^{-\frac{\alpha}{b}-is_1}\hat{B}^{-\frac{\alpha}{b}-is_2} \\
&\times e^{-\pi i (\alpha^2+b^2s_1^2)}\left(\begin{array}{c}-2\frac{\alpha}{b} \\ -\frac{\alpha}{b}-is_1 \end{array}\right)_b F_b(\alpha+ibs_1,\alpha+ibs_2,2\alpha;-\mathbf{z}^{-1})\text{qdet}^{\frac{Q}{2b}}, \nonumber
\end{align}
with $\mathbf{z} = q B\hat{B}\hat{A}^{-1} = \text{qdet}^{1/2} \, (\gamma e^{2\phi} \beta) \, \text{qdet}^{-1/2} = \gamma e^{2\phi} \beta$ is the hyperbolic element. Here the $b$-binomial coefficients are:\footnote{These arise in the $b$-binomial theorem. For $u,v$ \emph{positive} self-adjoint operators satisfying $uv=q^2vu$, we have:
\begin{equation}
(u+v)^{it} = \int_{\mathbb{R}}q^{\tau(t-\tau)} \left(\begin{array}{c}it \\ i\tau \end{array}\right)_b u^{it-i\tau}v^{i\tau} d\tau.
\end{equation}}
\begin{equation}
\left(\begin{array}{c}t \\ \tau \end{array}\right)_b = \frac{S_b(Q+bt)}{S_b(Q+b\tau)S_b(Q+bt-b\tau)}.
\end{equation}
The $q$-hypergeometric function $F_b$ is:
\begin{equation}
F_b(\alpha,\beta,\gamma;\mathbf{z}) \equiv \frac{S_b(\gamma)}{S_b(\alpha)S_b(\beta)} \int_{\mathcal{C}}d\tau (-\mathbf{z})^{ib^{-1}\tau}e^{-\pi\tau(\alpha+\beta-\gamma)}\frac{S_b(\alpha+i\tau)S_b(\beta+i\tau)S_b(-i\tau)}{S_b(\gamma+i\tau)}.
\end{equation}
Plugging these in, the explicit matrix element of \cite{Ip} becomes:
\begin{align}
\label{eq:Ip}
&R_{s_1s_2}^\alpha(g) = A^{-\frac{\alpha}{b}+is_1}B^{-\frac{\alpha}{b}-is_1}\hat{B}^{-\frac{\alpha}{b}-is_2} e^{-\pi i (\alpha^2+b^2s_1^2)} \\
&\hspace{-0.15cm}\times\frac{S_b(\alpha-ibs_1)}{S_b(\alpha+ibs_2)} \int_{\mathcal{C}}d\tau \mathbf{z}^{-i\frac{\tau}{b}}e^{-i\pi\tau(bs_1+bs_2)}\frac{S_b(\alpha+ibs_1+i\tau)S_b(\alpha+ibs_2+i\tau)S_b(-i\tau)}{S_b(2\alpha+i\tau)}\text{qdet}^{\frac{Q}{2b}}. \nonumber
\end{align}
Finally, using the identifications
\begin{align}
A \to e^\phi, \qquad B \to \gamma e^\phi, \qquad \hat{B} \to \beta,
\end{align}
we see that \eqref{eq:Ip} matches with \eqref{eq:toco}, up to the prefactor $\tilde{N}$ that does not depend on the coordinates $(\gamma,\phi,\beta)$.

\section{Some applications}
\label{s:appli}
In this section, we deduce some properties of the representation matrix elements, that are in part of direct relevance for gravitational calculations. These results are made technically possible by virtue of the explicit formula \eqref{eq:GEb}.

\subsection{Casimir difference equation(s)}
In the undeformed set-up, the $\mathfrak{sl}(2,\mathbb{R})$ Casimir operator is diagonalized by the irreducible representation matrix elements. Mixed parabolic representation matrix elements lead to a simplified Casimir equation which is just the Liouville equation.\footnote{The argument is well-known, see e.g. appendix F of \cite{Blommaert:2018oro} for a discussion in the JT gravity framework, or \cite{Gerasimov:1996zk}.} The $q$-analogue of the latter was shown in \cite{Kharchev:2001rs} to be satisfied by the Whittaker function \eqref{eq:whit}. As an application of our proposal \eqref{eq:GEb} and calculational procedure, we will here derive the Casimir eigenvalue equations that the hyperbolic $q$-representation matrix elements \eqref{eq:final} satisfy.

The modular double quantum algebra has two Casimir operators. The Casimir operator corresponding to $U_q(\mathfrak{sl}(2,\mathbb{R}))$ commutes with all generators $E,F,H$ and $\tilde{E},\tilde{F},\tilde{H}$. It is given by the expression (up to a choice of normalization):
\begin{equation}
\label{eq:cas}
\hat{\mathcal{C}} = \frac{q^{2H+1}+q^{-2H-1}}{(q-q^{-1})^2} + FE.
\end{equation}
There is also a dual Casimir operator coming from $U_{\tilde{q}}(\mathfrak{sl}(2,\mathbb{R}))$:
\begin{equation}
\hat{\tilde{\mathcal{C}}} = \frac{\tilde{q}^{2\tilde{H}+1}+\tilde{q}^{-2\tilde{H}-1}}{(\tilde{q}-\tilde{q}^{-1})^2} + \tilde{F}\tilde{E}.
\end{equation}
We now apply the Casimir operator $\hat{\mathcal{C}}$ within the representation matrix element as
\begin{equation}
\left\langle s_1 \right| g\, \hat{\mathcal{C}} \left|s_2\right\rangle.
\end{equation}
On the one hand, the Casimir is proportional to the unit matrix in the principal series representations (since it is irreducible), so we have by explicitly computing $\hat{\mathcal{C}}$ using \eqref{PT}:
\begin{equation}
\left\langle s_1 \right| g\, \hat{\mathcal{C}} \left|s_2\right\rangle  = \frac{\cosh 2 \pi b s}{2\sin^2 \pi b^2}\left\langle s_1 \right| g \left|s_2\right\rangle. 
\end{equation}
On the other hand, we can manipulate it into a difference operator that acts on the representation matrix as follows. 
We use \eqref{eq:main} once again and compute the LHS with the insertion of $\hat{\mathcal{C}}$: $\left\langle \nu \right| g\, \hat{\mathcal{C}} \left|\mu\right\rangle$. The desired hyperbolic representation matrix element then follows again by Laplace transforming as in \eqref{eq:inttrans}. We hence first evaluate $\left\langle \nu \right| g\, \hat{\mathcal{C}} \left|\mu\right\rangle$. In inserting the Casimir operator \eqref{eq:cas}, the term with the Cartan generator $q^{\pm(2H+1)}$ just contributes a linear combination of shift operators on the $\phi$-coordinate as:
\begin{equation}
\frac{qT^\phi_{i\pi b^2} + q^{-1} T^\phi_{-i\pi b^2}}{(q-q^{-1})^2}, \qquad T^\phi_a f(\phi) \equiv f(\phi+a).
\end{equation}
The $FE$ term in \eqref{eq:cas} gives after commuting the $F$ past $g$ and using that bra and ket diagonalize $F^\dagger$ and $E$ respectively as in \eqref{eq:whitv1} and \eqref{eq:whitv2}, an insertion of 
\begin{equation}
-\frac{\nu\mu}{(q-q^{-1})^2}.
\end{equation}
This term can be rewritten in terms of $q$-derivatives acting on the group element itself using the identities:
\begin{align}
\label{eq:qderi}
\left(\frac{d}{d\gamma}\right)_{\hspace{-0.1cm}q^{-2}} g_b(\gamma f) &= \frac{f}{q^{-1}-q} g_b(\gamma f), \\
\left(\frac{d}{d\beta}\right)_{\hspace{-0.1cm}q^2} g_b^*(\beta e) &= \frac{e}{q-q^{-1}}g_b^*(\beta e),
\end{align}
which are directly derived using \eqref{eq:eb}, \eqref{eq:eb2}, and represent the fact that $g_b$ is a $q$-exponential function.\footnote{Notice that this diverges when $b\to 0$, reflecting the fact that the dual quantum algebra contribution to the derivative diverges in this limit.} Here we have used the textbook $q$-derivative (for suitable choice of $q$), defined as
\begin{equation}
\left(\frac{d}{dx}\right)_{\hspace{-0.1cm}q} f(x) \equiv \frac{f(qx)-f(x)}{qx-x}.
\end{equation}
The integral transformations \eqref{eq:inttrans} can then be done immediately. Likewise, we have the dual identities:\footnote{The quantum dilogarithm $g_b$ has the unique property that both its $q^{-2}$-derivative and $\tilde{q}^{-2}$-derivative are proportional to $g_b$ itself, signaling a duality-invariant exponential function indeed.}
\begin{align}
\left(\frac{d}{d\gamma^{\frac{1}{b^2}}}\right)_{\hspace{-0.1cm}\tilde{q}^{-2}} g_b(\gamma f) &= \frac{\tilde{f}}{\tilde{q}^{-1}-\tilde{q}} g_b(\gamma f), \\
\left(\frac{d}{d\beta^{\frac{1}{b^2}}}\right)_{\hspace{-0.1cm}\tilde{q}^2} g_b^*(\beta e) &= \frac{\tilde{e}}{\tilde{q}-\tilde{q}^{-1}}g_b^*(\beta e),
\end{align}
that can be used to derive the dual Casimir equation. We end up with the Casimir difference equation and its dual:
\begin{align}
\label{eq:Casss}
&\hspace{-0.2cm}\Bigg(\frac{qT^\phi_{i\pi b^2} + q^{-1} T^\phi_{-i\pi b^2}}{(q-q^{-1})^2} + \left(\frac{d}{d\gamma}\right)_{\hspace{-0.1cm}q^{-2}} \hspace{-0.15cm}  e^{-2\phi} \left(\frac{d}{d\beta}\right)_{\hspace{-0.1cm}q^2} \Bigg) R_{s_1s_2}(g)  = \frac{\cosh 2 \pi b s}{2\sin^2 \pi b^2}R_{s_1s_2}(g), \\
&\hspace{-0.2cm}\Bigg(\frac{\tilde{q}T^\phi_{i\pi} + \tilde{q}^{-1} T^\phi_{-i\pi}}{(\tilde{q}-\tilde{q}^{-1})^2} + \left(\frac{d}{d\gamma^{\frac{1}{b^2}}}\right)_{\hspace{-0.1cm}\tilde{q}^{-2}} \hspace{-0.15cm} e^{-2\phi/b^2} \left(\frac{d}{d\beta^{\frac{1}{b^2}}}\right)_{\hspace{-0.1cm}\tilde{q}^2} \Bigg) R_{s_1s_2}(g)  = \frac{\cosh 2 \pi b^{-1} s}{2\sin^2 \pi b^{-2}}R_{s_1s_2}(g). \nonumber
\end{align}
Let us make some comments.
\begin{itemize}
\item
When using $b\to b^{-1}$ to find the second equation, one has to make sure to transform the coordinates $(\gamma,\phi,\beta)$ in the correct way. Alternatively, one can work with the invariant rescaled coordinates defined around \eqref{eq:rescrep}.
\item
Since the coordinates $(\gamma,\phi,\beta)$ are non-commutative, care has to be taken in the ordering of the different factors as written. In particular, we need to order the coordinates as $(\gamma,\phi,\beta)$ as done in the decomposition of $g$ \eqref{eq:GEb} and explicitly in $R_{s_1s_2}(g)$ as e.g. done in \eqref{eq:prefinal} and \eqref{eq:final}. Our way of writing these equations implies that the $\gamma$-derivative acts from the left, the $\beta$-derivative acts from the right, and the $e^{-2\phi}$ factor needs to be applied in the ``middle'' of $g$.
\item
It is instructive to explicitly check that \eqref{eq:final} satisfies these difference equations \eqref{eq:Casss}. For the second term of \eqref{eq:Casss}, one uses the properties:
\begin{align}
\left(\frac{d}{d\gamma}\right)_{\hspace{-0.1cm}q^{-2}} \gamma^{-is_1+i\zeta/b} &= \gamma^{-is_1+i\zeta/b-1}\frac{q^{2is_1-2i\zeta/b}-1}{q^{-2}-1}, \\
\left(\frac{d}{d\beta}\right)_{\hspace{-0.1cm}q^{2}}\beta^{-is_2+i\zeta/b} &= \beta^{-is_2+i\zeta/b-1}\frac{q^{-2is_2+2i\zeta/b}-1}{q^{2}-1},
\end{align}
such that the second term causes an effective shift $(\gamma e^{2\phi} \beta)^{i\zeta/b} \to (\gamma e^{2\phi} \beta)^{i\zeta/b-1}$ in the integrand of \eqref{eq:final}. Using then a contour shift $i\zeta \to i\zeta + b$ and the defining shift properties of the double sine function $S_b(x+b^{\pm 1}) = 2 \sin (\pi b^{\pm 1} x) S_b(x)$, one can explicitly show that \eqref{eq:Casss} is satisfied. The dual equation is checked analogously.
\item
In general, solving difference equations has an enormous ambiguity since the values of the unknown function are only related at discrete points. This is reflected in the presence of arbitrary periodic functions (sometimes called ``quasi-constants'') in the general solution. However, assuming $b^2$ is irrational, the pair of difference equations \eqref{eq:Casss} associated to the modular double leads to a ``dense'' covering of the $(\gamma,\phi,\beta)$ coordinate regions by combining back-and-forwards shifts of both $b$ and $b^{-1}$.
\end{itemize}

\subsection{Regular representation of the modular double quantum group}
The Casimir eigenvalue equation of an ordinary Lie group $G$ is the result of decomposing the regular representation into its irreducible components. Here we use \eqref{eq:GEb} to directly construct the regular representation of the modular double from first principles, and show that it indeed leads to the pair of Casimir equations \eqref{eq:Casss}.

The left-regular representation of any Lie group $G$ is defined by acting on the set of functions in $L^2(G)$ as:
\begin{equation}
f(h) \to f(g^{-1} \cdot h).
\end{equation}
Infinitesimally, this group action leads to a differential operator $\hat{L}_i$, defined by the relation
\begin{equation}
\hat{L}_i f(h) = \frac{d}{d\epsilon} f(e^{-\epsilon X_i} h)\vert_{\epsilon = 0},
\end{equation}
or:
\begin{equation}
\label{eq:lreg}
\hat{L}_i h = - X_i h.
\end{equation}
Analogously, one defines the right-regular realization as:
\begin{equation}
f(h) \to f(h \cdot g),
\end{equation}
leading to 
\begin{equation}
\label{eq:rreg}
\hat{R}_i g = g X_i.
\end{equation}
For quantum groups, we can directly work with \eqref{eq:lreg} and \eqref{eq:rreg} as defining the left- and right-regular realization in terms of difference operators $\hat{L}_i$ and $\hat{R}_i$.\footnote{It is not entirely clear how to start with one-parameter subgroups acting on $f(h)$ since $q$-exponentiating $F$ or $E$ does not lead to subgroups.}
For concreteness, we focus on \eqref{eq:rreg}, and collect the results on the left-regular realization at the end.

To extend this definition to the modular double of a quantum group, we use the ``doubled'' group element $g$ parametrized in \eqref{eq:GEb}. The quantum algebra generators are $(E,F,H)$ and $(\tilde{E},\tilde{F},\tilde{H})$ for the two copies. We will prove the following statement:
\begin{quote}
\emph{The regular representation of the modular double of the quantum group SL$_q(2,\mathbb{R})$, defined through either \eqref{eq:lreg} or \eqref{eq:rreg}, is equal to the modular double of the regular representation}.
\end{quote}

Let's start with the element $E$ in the right-regular realization \eqref{eq:rreg}. We want to find the operator $\hat{R}_E$ such that
\begin{equation}
\hat{R}_E (g_{b}(\gamma f) \, e^{2\phi H} \, g_{b}^*(\beta e)) = (g_{b}(\gamma f) \, e^{2\phi H} \, g_{b}^*(\beta e))\, E.
\end{equation}
Using \eqref{eq:qderi} it is immediate that
\begin{equation}
\label{eq:gen1}
\hat{R}_E = i\overleftarrow{\left(\frac{d}{d\beta}\right)_{\hspace{-0.1cm}q^2}}.
\end{equation}
This operator as written acts \emph{from the right} on any expression, which we depict by the arrow on top. Next let's look at the Cartan element $K = q^H$ in the form:
\begin{equation}
\hat{R}_K (g_{b}(\gamma f) \, e^{2\phi H} \, g_{b}^*(\beta e)) = (g_{b}(\gamma f) \, e^{2\phi H} \, g_{b}^*(\beta e))\, q^H,
\end{equation}
so that we read off:\footnote{We refrain from putting an arrow on $T^\phi$ since this does not matter when acting on $g$.}
\begin{equation}
\label{eq:gen2}
\hat{R}_K = T^\phi_{\log q/2} \overleftarrow{R^\beta_{1/q}},
\end{equation}
where we used
\begin{equation}
\label{eq:lem2}
g_{b}^*(\beta e) q^H = q^H g_{b}^*\left(\frac{\beta}{q} e\right),
\end{equation}
and defined the scaling operator $R^\beta_a f(\beta) \equiv f(a\beta)$. Finally, the hardest generator is $F$:
\begin{equation}
\hat{R}_F (g_{b}(\gamma f) \, e^{2\phi H} \, g_{b}^*(\beta e)) = (e_{q^{-2}}^{\gamma F} \, e^{2\phi H} \, g_{b}^*(\beta e))\, F.
\end{equation}
We use the property
\begin{align}
\label{eq:lem3}
g_{b}^*(\beta e) F 
&= F g_{b}^*(\beta e) - i\beta g_{b}^*(\beta e)\frac{q^{2H}}{q-q^{-1}}+i\beta g_{b}^*\left(\frac{\beta}{q^2} e\right) \frac{q^{-2H}}{q-q^{-1}},
\end{align}
and obtain\footnote{
As an example of how one works with expression such as these, we work out the first term in detail:
\small
\begin{align}
g_{b}(\gamma f) \, e^{2\phi H} \, g_{b}^*(\beta e) {\color{blue}(-i)T^\phi_{\log q}\overleftarrow{\left(\frac{d}{d\gamma}\right)_{\hspace{-0.1cm}q^{-2}}} \overleftarrow{R^\beta_{q^{-2}}} e^{-2\phi}} \nonumber
&= g_{b}(\gamma f) \, e^{2\phi H} q^{2H }\, g_{b}^*(\beta e) {\color{blue} (-i)\overleftarrow{\left(\frac{d}{d\gamma}\right)_{\hspace{-0.1cm}q^{-2}}} \overleftarrow{R^\beta_{q^{-2}}} e^{-2\phi}} \nonumber \\
&= g_{b}(\gamma f) \, (-i)\overleftarrow{\left(\frac{d}{d\gamma}\right)_{\hspace{-0.1cm}q^{-2}}} e^{2\phi H} \, g_{b}^*(\beta e) {\color{blue}\overleftarrow{R^\beta_{q^{-2}}} e^{-2\phi}} \nonumber \\
&= g_{b}(\gamma f) F\,  e^{2\phi H} \, g_{b}^*\left(\frac{\beta}{q^2} e \right) {\color{blue}e^{-2\phi}} \nonumber \\
&= g_{b}(\gamma f) F\, e^{-2\phi} e^{2\phi H} \, g_{b}^*(\beta e) \nonumber \\
&= g_{b}(\gamma f) \, e^{2\phi H} F \, g_{b}^*(\beta e)
\end{align}
\normalsize
}
\begin{equation}
\label{eq:gen3}
\hat{R}_F 
= -iT^\phi_{\log q}\overleftarrow{\left(\frac{d}{d\gamma}\right)_{\hspace{-0.1cm}q^{-2}}} \overleftarrow{R^\beta_{q^{-2}}} e^{-2\phi} -i \frac{\overleftarrow{R^\beta_{q^{-2}}}T^\phi_{\log q} -  T^\phi_{-\log q}}{q-q^{-1}} \beta.
\end{equation}
The Casimir operator can be evaluated and is of the form\footnote{Care has to be taken for the swapped ordering in which the operators are applied from the right of the expression. In particular, the $q$-derivative $\overleftarrow{\left(\frac{d}{d\beta}\right)_{\hspace{-0.1cm}q^{2}}}$ appears a priori on the left of the second term, but can be pulled through in a second step to match with the written expression.}
\begin{align}
&\hat{\mathcal{C}} = \frac{q\hat{R}_K^2+q^{-1}\hat{R}_K^{-2}}{(q-q^{-1})^2} + \hat{R}_F \hat{R}_E \\
&= \frac{qT^\phi_{\log q} + q^{-1} T^\phi_{-\log q}}{(q-q^{-1})^2} + T^\phi_{\log q}\overleftarrow{\left(\frac{d}{d\gamma}\right)_{\hspace{-0.1cm}q^{-2}}} R^\beta_{q^{-2}} e^{-2\phi} \overleftarrow{\left(\frac{d}{d\beta}\right)_{\hspace{-0.1cm}q^{2}}}. \nonumber
\end{align}
If one instead defines the operators such that they appear directly ordered in the correct place in the expression, as in the previous subsection, the expression could be written as
\begin{equation}
\hat{\mathcal{C}} = \frac{qT^\phi_{\log q} + q^{-1} T^\phi_{-\log q}}{(q-q^{-1})^2} + \overrightarrow{\left(\frac{d}{d\gamma}\right)_{\hspace{-0.1cm}q^{-2}}} e^{-2\phi} \overleftarrow{\left(\frac{d}{d\beta}\right)_{\hspace{-0.1cm}q^{2}}},\label{142}
\end{equation}
which precisely matches with the first equation in \eqref{eq:Casss}.

Analogously, one can work out the dual generators. E.g.
\begin{equation}
\hat{R}_{\tilde{E}} (g_{b}(\gamma f) \, e^{2\phi H} \, g_{b}^*(\beta e)) = g_{b}(\gamma f) \, e^{2\phi H} \, g_{b}^*(\beta e) {\tilde{E}},
\end{equation}
which leads to
\begin{equation}
\label{eq:gen4}
\hat{R}_{\tilde{E}} = i\overleftarrow{\left(\frac{d}{d\beta^{1/b^2}}\right)_{\hspace{-0.1cm}\tilde{q}^2}}.
\end{equation}
Analogously we have:
\begin{equation}
\label{eq:gen5}
\hat{R}_{\tilde{K}} = (\hat{R}_{K})^{1/b^2} = T^\phi_{\pi i/2} \overleftarrow{R^\beta_{e^{-\pi i}}}.
\end{equation}
For the last one, we need now the dual identity:
\begin{align}
\label{eq:lem4}
g_{b}^*(\beta e) \tilde{F} 
&= \tilde{F} g_{b}^*(\beta e) - i\beta^{1/b^2} g_{b}^*(\beta e)\frac{\tilde{q}^{2\tilde{H}}}{\tilde{q}-\tilde{q}^{-1}}+i\beta^{1/b^2} g_{b}^*\left(\frac{\beta}{\tilde{q}^{2b^2}} e\right) \frac{\tilde{q}^{-2\tilde{H}}}{\tilde{q}-\tilde{q}^{-1}},
\end{align}
which finally leads to the expression:
\begin{equation}
\label{eq:gen6}
\hat{R}_{\tilde{F}} 
= -iT^\phi_{\pi i}\overleftarrow{\left(\frac{d}{d\gamma^{1/b^2}}\right)_{\hspace{-0.1cm}\tilde{q}^{-2}}} \overleftarrow{R^\beta_{e^{-2\pi i}}} e^{-2\phi/b^2} -i \frac{\overleftarrow{R^\beta_{e^{-\pi i}}}T^\phi_{\pi i} -  T^\phi_{-\pi i}}{\tilde{q}-\tilde{q}^{-1}} \beta.
\end{equation}
This results in the dual Casimir on the second line of \eqref{eq:Casss}. \\

The Casimir operator in the regular representation of the algebra is interpreted as the analogue of the Laplacian on the quantum group manifold. In the case of the modular double quantum group, there are two Casimir operators that are $b\to 1/b$ dual to each other. Representation matrix elements are simultaneous eigenfunctions of both, as we have explicitly checked for the hyperbolic representation matrix element in equation \eqref{eq:Casss}, but can also be explicitly seen in the simpler case of the Whittaker function $R_{\nu\mu}(\phi)$ \eqref{eq:whit} by checking explicitly that it satisfies:
\begin{align}
\label{eq:caswhit}
\Bigg(\frac{qT^\phi_{i\pi b^2} + q^{-1} T^\phi_{-i\pi b^2}}{(q-q^{-1})^2} + \frac{\mu \nu}{4 \sin^2 \pi b^2} e^{-2\phi}  \Bigg) R_{\nu\mu}(\phi) &= \frac{\cosh 2 \pi b s}{2\sin^2 \pi b^2} R_{\nu\mu}(\phi), \\
\Bigg(\frac{\tilde{q}T^\phi_{i\pi} + \tilde{q}^{-1} T^\phi_{-i\pi}}{(\tilde{q}-\tilde{q}^{-1})^2} + \frac{\mu^{1/b^2}\nu^{1/b^2}}{4 \sin^2 \pi b^{-2}} e^{-2\phi/b^2}\Bigg) R_{\nu\mu}(\phi)  &= \frac{\cosh 2 \pi b^{-1} s}{2\sin^2 \pi b^{-2}} R_{\nu\mu}(\phi). \nonumber
\end{align}

The three generators \eqref{eq:gen1}, \eqref{eq:gen2} and \eqref{eq:gen3} are the same as those found by using the SL$_q(2)$ Gauss-Euler decomposition \eqref{UqGE}, as we computed explicitly in \cite{Blommaert:2023opb}.\footnote{This is up to factors of $i$ for the parabolic generators taking instead $-i$\eqref{eq:gen1} and $i$\eqref{eq:gen3}. These factors of $i$ are expected as explained in section \ref{s:constr}.} Taking their generator duals $b\to 1/b$ (and being careful about the scaling of the coordinates $(\gamma,\phi,\beta$) as before), we immediately obtain \eqref{eq:gen4}, \eqref{eq:gen5} and \eqref{eq:gen6} without any calculation. \textbf{We hence conclude that the modular double of the regular representation equals the regular representation of the modular double, as defined and determined in this subsection.} \\

For completeness, we collect the generators of the left-regular realization \eqref{eq:lreg}:
\begin{align}
\hat{L}_F &= i\left(\frac{d}{d\gamma}\right)_{\hspace{-0.1cm}q^{-2}}, \\
\hat{L}_K &= T^\phi_{-\log q/2} R^\gamma_{q}, \\
\hat{L}_E &= -ie^{-2\phi} R^\gamma_{q^{2}} \left(\frac{d}{d\beta}\right)_{\hspace{-0.1cm}q^{2}} T^\phi_{-\log q} - i\gamma \frac{T^\phi_{\log q}- R^\gamma_{q^2}T^\phi_{-\log q}}{q-q^{-1}}, \\
\hat{L}_{\tilde{F}} &= i\left(\frac{d}{d\gamma^{1/b^2}}\right)_{\hspace{-0.1cm}q^{-2}}, \\
\hat{L}_{\tilde{K}} &= (\hat{L}_{K})^{1/b^2} = -T^\phi_{-\pi i/2} R^\gamma_{e^{\pi i}}, \\
\hat{L}_{\tilde{E}} &= -ie^{-2\phi/b^2} R^\gamma_{e^{2\pi i}} \left(\frac{d}{d\beta^{1/b^2}}\right)_{\hspace{-0.1cm}q^{2}} T^\phi_{-\pi i} - i\gamma \frac{T^\phi_{\pi i}- R^\gamma_{e^{2\pi i}}T^\phi_{-\pi i}}{\tilde{q}-\tilde{q}^{-1}},
\end{align}
where we used the identities:
\begin{equation}
g_{b}\left(\frac{\gamma}{q} f\right) q^H = q^H g_{b}(\gamma f),
\end{equation}
\begin{equation}
E g_{b}(\gamma f) = g_{b}(\gamma f) E + i\gamma \frac{q^{2H}}{q-q^{-1}}g_{b}(q^2\gamma f) - i\gamma \frac{q^{-2H}}{q-q^{-1}} g_{b}(\gamma f).
\end{equation}
This leads to the same Casimir operator \eqref{142} and its modular dual.

\subsection{One-sided wavefunctions and gravitational interpretation}
\label{sec:onesided}
As a final application, we write down an expression for the representation matrix with on the left a hyperbolic eigenstate, and on the right a generalized parabolic eigenstate as follows. The right boundary state is generalized into the one-parameter family of states \cite{Kharchev:2001rs}:
\begin{equation}
E \phi_{\mu,\alpha_2}(t) = \frac{i \mu}{q-q^{-1}} q^{2\alpha_2 H}\phi_{\mu,\alpha_2}(t), \qquad \tilde{E} \phi_{\mu,\alpha_2}(t) = \frac{i \mu^{1/b^2}}{\tilde{q}-\tilde{q}^{-1}} \tilde{q}^{2\alpha_2 \tilde{H}}\phi_{\mu,\alpha_2}(t).
\end{equation}
Setting $\alpha_2=0$ reduces the right boundary state to the one studied before, but we choose to be slightly more general here. Going through an analogous computation as before where we set $\phi_{\mu,\alpha_2}(t) \equiv \left\langle t\right|\left. \mu,\alpha_2 \right\rangle$ and use 
\begin{align}
\left\langle \nu \right| g \left|\mu,\alpha_2\right\rangle = \int_{-\infty}^{+\infty} ds_1 \, \left\langle \nu \right| \left. s_1\right\rangle \left\langle s_1 \right| g \left|\mu,\alpha_2\right\rangle,
\end{align}
we arrive at the single-sided representation matrix:
\begin{align}
\label{eq:1s}
\left\langle s_1 \right| g \left|\mu,\alpha_2\right\rangle \equiv \,\, &R_{s_1;\mu\alpha_2}(g) =N \gamma^{-is_1} \int_{\mathcal{C}} d\zeta\, e^{\pi i (\alpha_2-\frac{1}{2})\zeta^2 + ((\alpha_2+\frac{1}{2}) \pi Q +\pi i b s_1)\zeta}\mu^{-i\frac{\zeta}{b}} \\
&\times\frac{S_b(ibs_1-i\zeta)S_b(i\zeta+\alpha) S_b(i\zeta+\bar{\alpha})}{S_b(ibs_1 + \bar{\alpha})} (\gamma e^{2 \phi})^{i \frac{\zeta}{b}} g_b^*(\beta \mu e^{-2\pi b \alpha_2 \zeta}), \nonumber
\end{align}
with the explicit normalization factor
\begin{equation}
N = b^{1/2} e^{-\frac{\pi}{2}s_1}e^{-\frac{\pi i b^2}{2}s_1^2}e^{-\frac{\pi b^2}{2}s_1}e^{-\pi i \alpha_2\left(\frac{Q^2}{4}+\mathfrak{s}^2\right)}\mu^{-\frac{\alpha}{b}}.
\end{equation}
The choice $\alpha_2 = \pm \frac{1}{2}$ is prefered in the context of Liouville gravity, but we leave it arbitrary here. Moreover, for physics applications one might want to set $\beta=0$ and consider the quantum subgroup generated by $\gamma,\phi$ only, corresponding to an intermediate case between the Whittaker function only depending on $\phi$, and the full irrep matrix element depending on all three coordinates $\gamma,\phi,\beta$. Concretely, this just means one deletes the last factor of $g_b^*(.)$ in \eqref{eq:1s} since $g_b(0)=1$. 

Such one-sided wavefunctions are of interest when describing the Hilbert space exterior to a black hole as follows. Starting with a two-sided Hilbert space (and wavefunction), one can attempt to split the Hilbert space into a left (L) piece and a right (R) piece. However, in gauge theories and gravity alike, such a splitting cannot be done directly due to the non-local constraints acting on physical states in the Hilbert space \cite{Buividovich:2008gq,Casini:2013rba,Donnelly:2014gva,Donnelly:2016auv}. Factorization can be achieved by enlarging the Hilbert space and allowing surface charges at the splitting (or entangling) surface. In lower-dimensional gauge theory models, this is done explicitly by factorizing using the defining property of a representation:
\begin{equation}
R_{ab}(g_1g_2) = \sum_c R_{ac}(g_1) R_{cb}(g_2), \qquad g_1,g_2 \in G,\quad a,b,c = 1 \hdots \text{dim }R,
\end{equation}
where the gravitational wavefunction $\psi_{ab}(g) \sim R_{ab}(g)$. This factorizes a two-sided wavefunction into the product of wavefunctions with the index $c$ living at the splitting surface. The splitting surface itself is also the entangling surface or a black hole horizon according to an observer whose observations are restricted to a single side. Exploiting the fact that lower-dimensional gravitational models have a gauge theoretic description, we investigated a similar factorization in several gravity models in earlier work \cite{Blommaert:2018iqz,Mertens:2022ujr}. It turned out that in both these cases, the splitting index $c$ has to be a hyperbolic index $s$ as discussed here. Hence in the case when the underlying group theoretic structure is the modular double U$_q(\mathfrak{sl}(2,\mathbb{R})) \otimes $U$_{\tilde{q}}(\mathfrak{sl}(2,\mathbb{R}))$, the above one-sided wavefunctions \eqref{eq:1s} are precisely these split gravitational wavefunctions, with one asymptotic index and one index on the entangling surface. Moreover, the hyperbolic index $s$ is then labeling edge state degrees of freedom that are inaccessible for an outside fiducial or one-sided observer, an observer whose observations are restricted to information living on just this one side. The set of all possibilities for $s$ describes the different black hole microstates that are consistent with its macroscopic properties (i.e. its total mass), encoded in the Casimir eigenvalue. This is the picture we advocated for in \cite{Blommaert:2018iqz,Mertens:2022ujr}.

\begin{figure}[!htb]
\centering
\includegraphics[width=0.75\textwidth]{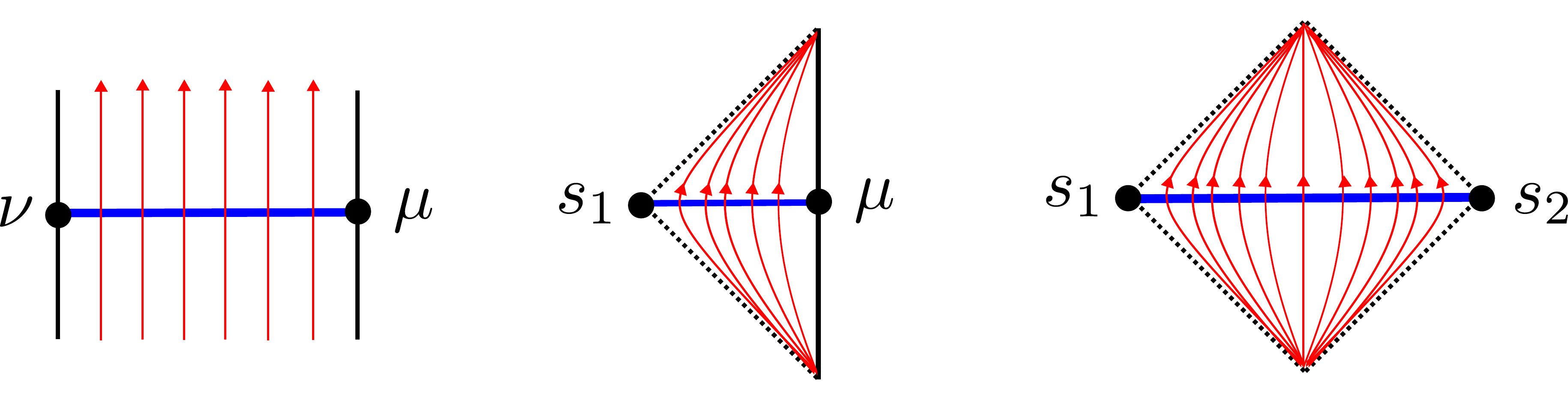}
\caption{Different gravitational wavefunctions (blue) and the modular flow (red) relevant for each of them in the gravitational application. Dashed diagonal lines are black hole horizons. Hyperbolic $s$-labels are attributed to fixed points of the modular flow (black hole horizons), and count black hole microstates. Left: two-sided wavefunction (two holographic boundaries) \eqref{eq:whit}. Middle: one-sided wavefunction (one holographic boundary) \eqref{eq:1s}. Right: Interior wavefunction (no holographic boundaries) \eqref{eq:final}.}
\label{mflow}
\end{figure}

We summarize the gravitational interpretation of these different gravitational wavefunctions in figure \ref{mflow}.

\section{Concluding Remarks}
We have provided evidence that matrix exponentiation to go from the quantum algebra to the quantum group can be shown to work (in the sense of \eqref{eq:introfinal}) for the modular double U$_q(\mathfrak{sl}(2,\mathbb{R})) \otimes $U$_{\tilde{q}}(\mathfrak{sl}(2,\mathbb{R}))$. This is particularly useful since this is how concrete calculations of BF amplitudes are done when there are boundary conditions that restrict some of the generators as discussed in the Introduction. Moreover, this calculational technique allows us to relate the two previously known representation matrix elements of SL$_q^+(2,\mathbb{R})$: Ip's hyperbolic representation matrix element \cite{Ip}, written in our parametrization in \eqref{eq:final}, and the Whittaker function \eqref{eq:whit} \cite{Kharchev:2001rs}. We applied our proposal to show that the representation matrix elements are simultaneous solutions to two Casimir eigenvalue equations \eqref{eq:Casss}, and we constructed a representation matrix element that mixes a hyperbolic and parabolic index \eqref{eq:1s}, relevant to describe one-sided wavefunctions in lower-dimensional gravity models. Structurally, we have constructed the regular representation of the modular double, reproducing both Casimir operators, and showed that it is equal to the modular double of the regular representation. More abstractly, in Appendix \ref{app:hopf} we have embedded and interpreted our proposal in terms of Hopf duality between the modular double quantum algebra and the (Hopf) dual matrix quantum group.

Moreover, the presented technique looks amenable to supersymmetrization to find the $\mathcal{N}=1$ hyperbolic representation matrix element starting with the known Whittaker function recently determined \cite{Fan:2021bwt}. These would be of importance for amplitudes in 2d $\mathcal{N}=1$ Liouville supergravity and 3d $\mathcal{N}=1$ supergravity.

\section*{Acknowledgments}
We thank A. Blommaert, Y. Fan, J. Sim\'on, G. Wong and S. Yao for discussions and collaborations related to this work.
TM acknowledges financial support from the European Research Council (grant BHHQG-101040024). Funded by the European Union. Views and opinions expressed are however those of the author(s) only and do not necessarily reflect those of the European Union or the European Research Council. Neither the European Union nor the granting authority can be held responsible for them.

\appendix

\section{Hopf algebra structure and duality}
\label{app:hopf}
In this appendix, we develop some of the underlying Hopf algebra of the modular double in the explicit Gauss-Euler coordinatization discussed in the main text. Our goal is to prove that the proposed formulas \eqref{eq:GEb} and \eqref{eq:expge} implement Hopf duality, generalizing how Lie algebras and Lie groups are related to the current case of a modular doubled quantum group.

\subsection{The coordinate Hopf algebra of the modular double}
The coordinate algebra SL$_q(2)$ is generated by the non-commutative variables $A,B,C,D$, with the co-product structure following from the matrix multiplication of the SL$(2,\mathbb{R})$ matrices
\begin{equation}
\begin{pmatrix}A & B 
    \\ C& D\end{pmatrix},
\end{equation}
leading to the co-product:
\begin{equation}
\begin{aligned}
    & \Delta(A)=A\otimes A' + B \otimes C',\quad \Delta(B)= A\otimes B'+B\otimes D' ,
    \\& \Delta(C)=C\otimes A' + D \otimes C' ,\quad \Delta(D)=C\otimes B' + D \otimes D',
\end{aligned}\label{coprod}
\end{equation}
where the prime denotes the matrix entries of the second matrix. In terms of the Gauss variables $(\gamma,\phi,\beta)$ where
\begin{equation}
\begin{pmatrix}A & B 
    \\ C& D\end{pmatrix} = \begin{pmatrix}e^\phi & e^\phi \beta \\ \gamma e^\phi & e^{-\phi}+\gamma e^\phi \beta\end{pmatrix},
\end{equation}
we write the co-product as:
\begin{align}
\label{eq:cop1}
\Delta(\gamma) &= \gamma + e^{-\phi} \gamma'(1+\beta\gamma')^{-1} e^{-\phi}, \\
\Delta(e^\phi) &= e^\phi(1+\beta\gamma')e^{\phi'}, \\
\Delta(\beta) &= \beta' + e^{-\phi'}(1+\beta \gamma')^{-1} \beta e^{-\phi'},
\end{align}
where we dropped the tensor product $\otimes$, adding the convention that all primed variables commute with the unprimed variables. One can furthermore transfer the co-product on $e^\phi$ to one for $\phi$ as (see eq. (58)-(59) of \cite{Fronsdal:1991gf} for the technical argument):\footnote{In the $q\to 1$ limit, the last sum reduces to the series expansion for $\ln(1+\beta\gamma')$.}
\begin{equation}
\Delta(\phi) = \phi + \phi' - \pi b^2 \sum_{n=1}^{+\infty}\frac{(-\beta\gamma')^n}{\sin (\pi b^2 n)}.
\end{equation}
The co-unit in these variables is given by
\begin{equation}
\epsilon(\phi) = 0, \quad \epsilon(\beta) = 0, \quad \epsilon(\gamma) = 0.
\end{equation}
Finally, the antipode $S(.)$ is given by
\begin{align}
S(e^\phi) = e^{-\phi} + \gamma e^\phi \beta, \quad S(e^\phi \beta) = -\beta e^\phi, \quad S(\gamma e^\phi) = - e^\phi \gamma.
\end{align}

For the modular double, we have additionally a dual $b\to 1/b$ set of variables. The dual triple ($\gamma^{\frac{1}{b^2}}, \phi/b^2, \beta^{\frac{1}{b^2}}$) are non-commutative variables satisfying the relations in terms of the dual deformation parameter $\tilde{q}$:
\begin{equation}
\label{eq:comma}
e^{\frac{\phi}{b^2}}\gamma^{\frac{1}{b^2}} = \tilde{q} \gamma^{\frac{1}{b^2}} e^{\frac{\phi}{b^2}}, \quad e^{\frac{\phi}{b^2}} \beta^{\frac{1}{b^2}}= \tilde{q} \beta^{\frac{1}{b^2}} e^{\frac{\phi}{b^2}}, \quad [\beta^{\frac{1}{b^2}},\gamma^{\frac{1}{b^2}}]=0,
\end{equation}
and the analogous co-product:
\begin{align}
\label{eq:cod1}
\Delta(\gamma^{\frac{1}{b^2}}) &= \gamma^{\frac{1}{b^2}} + e^{-\frac{\phi}{b^2}} \gamma'^{\frac{1}{b^2}}(1+\beta^{\frac{1}{b^2}}\gamma'^{\frac{1}{b^2}})^{-1} e^{-\frac{\phi}{b^2}}, \\
\Delta(e^{\frac{\phi}{b^2}}) &= e^{\frac{\phi}{b^2}}(1+\beta^{1/b^2}\gamma'^{\frac{1}{b^2}})e^{\frac{\phi'}{b^2}}, \\
\Delta(\beta^{\frac{1}{b^2}}) &= \beta'^{\frac{1}{b^2}} + e^{-\frac{\phi'}{b^2}}(1+\beta^{\frac{1}{b^2}} \gamma'^{\frac{1}{b^2}})^{-1} \beta^{\frac{1}{b^2}} e^{-\frac{\phi'}{b^2}}.
\end{align}
The highly non-trivial feature is that this co-product of the dual variables is compatible with the definition of these dual variables as powers (or rescalings) of the original variables $(\gamma,\phi,\beta)$ as follows. We need to assume a positivity notion at this point for the operators $\gamma,\beta \geq 0$ and $\gamma',\beta' \geq 0$ in order to make sense of the operations that follow. Let us take the first co-product \eqref{eq:cop1} and take the $1/b^2$ power:
\begin{equation}
\label{eq:stadual}
\Delta(\gamma)^{\frac{1}{b^2}} = \left(\gamma + e^{-\phi} \gamma'(1+\beta\gamma')^{-1} e^{-\phi} \right)^{\frac{1}{b^2}}.
\end{equation}
To proceed, we repeatedly utilize the following lemma. If $u$ and $v$ form a Weyl pair, i.e. non-commutative positive variables satisfying $uv = q^2 vu$ with $q=e^{\pi i b^2}$, then one has \cite{Bytsko:2002br}:
\begin{equation}
\label{eq:weyl}
(u+v)^{\frac{1}{b^2}} = u^{\frac{1}{b^2}} + v^{\frac{1}{b^2}}.
\end{equation}
The term in the bracket on the RHS of \eqref{eq:stadual} is a sum of two Weyl variables. Hence:
\begin{equation}
\label{eq:sta2}
\Delta(\gamma)^{\frac{1}{b^2}} = \gamma^{\frac{1}{b^2}} + \left(e^{-\phi} \gamma'(1+\beta\gamma')^{-1} e^{-\phi} \right)^{\frac{1}{b^2}}.
\end{equation}
Denote the quantity in brackets in the second term as $I = e^{-\phi} \gamma'(1+\beta\gamma')^{-1} e^{-\phi}$. We note that its left-right inverse $I^{-1}$ is 
\begin{equation}
I^{-1} = e^\phi (1+\beta\gamma')\gamma'^{-1}e^\phi.
\end{equation}
Now take the $1/b^2$ power of the relation $I \cdot I^{-1} = 1 = I^{-1}\cdot I$. The inverse $I^{-1}$ is again a sum of two Weyl variables, hence:
\begin{equation}
(I^{-1})^{\frac{1}{b^2}} = e^{\frac{\phi}{b^2}} (1+\beta^{\frac{1}{b^2}}\gamma'^{\frac{1}{b^2}})\gamma'^{-\frac{1}{b^2}}e^{\frac{\phi}{b^2}}.
\end{equation}
Finally, this means the quantity we want $I^{1/b^2}$ is just the inverse of this:
\begin{equation}
I^{\frac{1}{b^2}} = e^{-\frac{\phi}{b^2}} \gamma'^{\frac{1}{b^2}}(1+\beta^{\frac{1}{b^2}}\gamma'^{\frac{1}{b^2}})^{-1}e^{-\frac{\phi}{b^2}}.
\end{equation}
Inserting this in \eqref{eq:sta2}, we find \eqref{eq:cod1}. Similarly, one can compute
\begin{equation}
\Delta(e^{\phi})^{\frac{1}{b^2}} = \left(e^\phi e^{\phi'} + e^\phi \beta \gamma' e^{\phi'}\right)^{\frac{1}{b^2}} = e^{\frac{\phi}{b^2}}(1+\beta^{\frac{1}{b^2}}\gamma^{\frac{1}{b^2}}) e^{\frac{\phi'}{b^2}} = \Delta(e^{\frac{\phi}{b^2}}),
\end{equation}
using again the Weyl pair property. Analogously the relation for $\Delta(\beta)^{\frac{1}{b^2}}$ can be derived. The co-unit of the dual variables is again compatible:
\begin{equation}
\epsilon(\phi/b^2) = 0, \quad \epsilon(\beta^{\frac{1}{b^2}}) = 0, \quad \epsilon(\gamma^{\frac{1}{b^2}}) = 0,
\end{equation}
and the same holds for the antipode, using the same Weyl pair lemma \eqref{eq:weyl} several times. \\

For the following, we denote the dual variables as $\tilde{\gamma} \equiv \gamma^{1/b^2}$ and $\tilde{\beta} \equiv \beta^{1/b^2}$. The above compatibility of the dual relations, i.e. following from taking suitable powers (or scalings) of the original variables $(\gamma,\phi,\beta)$, implies the following. The co-product on the modular double coordinate Hopf algebra is determined by matrix multiplication of the SL$(2,\mathbb{R})$ matrix in terms of ($\gamma,\phi,\beta$), where the \textbf{co-product on the dual variables is \emph{induced} from that of the direct variables $\gamma,\phi,\beta$}. We summarize all five co-product relations:
\begin{align}
\label{eq:co1}
\Delta(\tilde{\gamma}) &= \tilde{\gamma} + e^{-\phi/b^2} \tilde{\gamma}'(1+\tilde{\beta}\tilde{\gamma}')^{-1} e^{-\phi/b^2}, \\
\Delta(\gamma) &= \gamma + e^{-\phi} \gamma'(1+\beta\gamma')^{-1} e^{-\phi}, \\
\Delta(\phi) &= \phi + \phi' - \pi b^2 \sum_{n=1}^{+\infty}\frac{(-\beta\gamma')^n}{\sin (\pi b^2 n)} = \phi + \phi' - \pi \sum_{n=1}^{+\infty}\frac{(-\tilde{\beta}\tilde{\gamma}')^n}{\sin (\pi b^{-2} n)}, \\
\Delta(\beta) &= \beta' + e^{-\phi'}(1+\beta \gamma')^{-1} \beta e^{-\phi'}, \\
\label{eq:co2}
\Delta(\tilde{\beta}) &= \tilde{\beta}' + e^{-\phi'/b^2}(1+\tilde{\beta} \tilde{\gamma}')^{-1} \tilde{\beta} e^{-\phi'/b^2}
\end{align}
where the middle relation combines both the co-product for $\phi$ and $\phi/b^2$, which are proportional.

Now we can define in more detail the coordinate Hopf algebra for the modular double of SL$_q(2,\mathbb{R})$. It is generated by the ordered monomial basis elements: 
\begin{equation}
X^{mnopq} \equiv \tilde{\gamma}^{m}\gamma^{n} \phi^{o} \beta^{p} \tilde{\beta}^{q}, \qquad m,n,o,p,q \in \mathbb{N},
\end{equation}
where the dual element $\phi/b^2$ does not appear separately (since it is just a multiple of $\phi$). The dual elements $\tilde{\gamma}$ and $\tilde{\beta}$ do appear since they are non-polynomials powers of $\gamma$ and $\beta$ and can hence not be obtained in the basis of $\gamma^n \beta^p$. A general element $a$ in the coordinate algebra is then a linear combination of these basis elements:
\begin{equation}
a = \sum_{m,n,o,p,q = 0}^{+\infty} a_{mnopq}\, X^{mnopq}.
\end{equation}
The co-product on the basis elements can be expanded generally as
\begin{equation}
\label{eq:defcop}
\Delta(X^{abcde}) = \Delta(\tilde{\gamma})^m \Delta(\gamma)^n \Delta(\phi)^o \Delta(\beta)^p \Delta(\tilde{\beta})^q = \sum F^{abcde}_{mnopq \vert tuvwx}X^{mnopq} X'^{tuvwx},
\end{equation}
where we use the primed notation as earlier as a substitute for the tensor product notation. The explicit values of these structure coefficients can be deduced from \eqref{eq:co1}-\eqref{eq:co2} as we will do further on.

\subsection{Hopf duality}
Hopf duality of two Hopf algebras, with basis denoted as $P_\alpha$ and $X^\alpha$, is defined \cite{Klimyk:1997eb} by a bilinear mapping $\left\langle\, .\,,\,.\, \right\rangle$ satisfying $\langle P_\alpha,X^\beta \rangle = \delta_\alpha^\beta$ and the duality properties:
\begin{equation}
\label{eq:dual}
\langle \Delta(P_\alpha),X^a \otimes X^b\rangle = \langle P_\alpha, X^aX^b\rangle, \qquad \langle P_\alpha P_\beta, X^a\rangle = \langle P_a \otimes P_b, \Delta(X^a) \rangle,
\end{equation}
This means multiplication and co-multiplication of both Hopf algebras get interchanged under duality. It is indeed simple to show that the general expansion of the product and co-product of both bases have related expansion coefficients as:
\begin{alignat}{3}
\label{eq:exp1}
\Delta(P_\gamma) &= \sum_{\alpha,\beta} E^{\alpha\beta}_\gamma P_\alpha \otimes P_\beta, \qquad &&P_\alpha P_\beta && = \sum_{\gamma} F_{\alpha\beta}^\gamma P_\gamma,\\
\label{eq:exp2}
X^\alpha X^\beta &= \sum_c E^{\alpha\beta}_\gamma X^\gamma, \qquad
&&\Delta(X^\gamma) &&= \sum_{\alpha,\beta} F^{\gamma}_{\alpha\beta} X^\alpha \otimes X^\beta,
\end{alignat}
for structure coefficients $E^{\alpha\beta}_\gamma, F_{\alpha\beta}^\gamma$.

One can then show that the algebra bilinear
\begin{equation}
g \equiv \sum_\alpha X^\alpha P_\alpha,
\end{equation}
where the basis and its dual are summed over in a diagonal combination, forms a representation of the dual quantum matrix group \cite{Fronsdal:1991gf} (we recently reviewed this argument in appendix A of \cite{Blommaert:2023opb}). Indeed, define two algebra bilinears $g_1,g_2$ as:
\begin{equation}
g_1 \equiv \sum_\alpha X^\alpha P_\alpha, \qquad g_2 \equiv \sum_\alpha X'^\alpha P_\alpha,
\end{equation}
with different coordinates $X^\alpha$ and $X'^\alpha$. The product $g_1 \, g_2$ can then be explicitly expanded as
\begin{equation}
\label{eq:repgrou}
g_1 \, g_2 = \sum_{\alpha,\beta} X^\alpha X'^\beta P_\alpha P_\beta = \sum_{\gamma} \Delta(X^\gamma) P_\gamma, 
\end{equation}
The coordinates (i.e. the $X$'s) of the product matrix $g_1\,g_2$ are hence just the coproduct $\Delta(X^\gamma)$ which was constructed initially to match with the matrix multiplication. Hence the quantum group elements $g$ form a (co-)representation of the matrix quantum group.

As mentioned, the dual Hopf algebra is determined by having the product and co-product swapped \cite{Klimyk:1997eb}. This means for our case that the dual basis, which we will denote as $P_{tuvwx}$ has the \emph{product} of the basis elements as
\begin{equation}
P_{mnopq}P_{tuvwx} = \sum F^{abcde}_{mnopq \vert tuvwx} P_{abcde},
\end{equation}
with precisely the same structure coefficients as those appearing in the co-product in \eqref{eq:defcop}.

The technical step still to do is to explicitly construct the basis $P_{tuvwx}$ of the dual algebra. We follow the notation and strategy of \cite{Fronsdal:1991gf}.
From the explicit co-product expressions \eqref{eq:co1}-\eqref{eq:co2}, we immediately deduce the specific cases:
\begin{align}
F^{abcde}_{00000 \vert tuvwx} &= \delta^{a}_{t}\delta^{b}_u \delta^c_v \delta^d_w \delta^e_x, \\
F^{abcde}_{mnopq \vert 00000} &= \delta^{a}_{m}\delta^{b}_n \delta^c_o \delta^d_{p} \delta^e_q.
\end{align}
From this, we have $P_{00000}P_{abcde} = P_{abcde}P_{00000} = P_{abcde}$ so $P_{00000}=1$. We also immediately find:
\begin{align}
\label{eq:re1}
F^{abcde}_{10000 \vert tuvwx} &= [a]_{\tilde{q}^{-2}} \, \delta^{a}_{t+1}\delta^{b}_u \delta^c_v \delta^d_w \delta^e_x, \\
F^{abcde}_{01000 \vert tuvwx} &= [b]_{q^{-2}} \, \delta^{a}_{t}\delta^{b}_{u+1} \delta^c_v \delta^d_w \delta^e_x, \\
F^{abcde}_{mnopq \vert 00010} &= [d]_{q^{2}} \, \delta^{a}_{m}\delta^{b}_n \delta^c_o \delta^d_{p+1} \delta^e_q, \\
\label{eq:re2}
F^{abcde}_{mnopq \vert 00001} &= [e]_{\tilde{q}^{2}} \, \delta^{a}_{m}\delta^{b}_n \delta^c_o \delta^d_p \delta^e_{q+1}.
\end{align}
Labeling the five ``constituent'' basis elements of the dual algebra as:
\begin{align}
\label{eq:ini}
P_{10000} = i\tilde{F}, \, P_{01000} = iF, \, P_{00100} = 2H, \, P_{00010} = -iE, \, P_{00001} = -i\tilde{E},
\end{align}
and using the above relations \eqref{eq:re1}-\eqref{eq:re2}, we can decrease the first, second, fourth and fifth index of $P_{tuvwx}$ recursively. E.g. for the first index we can use:
\begin{equation}
P_{10000}P_{(t-1)uvwx} = [t]_{\tilde{q}^{-2}}P_{tuvwx},
\end{equation}
to obtain
\begin{equation}
P_{tuvwx} = P_{00v00} \, \frac{P_{10000}^t}{[t]_{\tilde{q}^{-2}}!}\frac{P^u_{01000}}{[u]_{q^{-2}}!} \frac{P^w_{00010}}{[w]_{q^{2}}!} \frac{P^x_{00001}}{[x]_{\tilde{q}^{2}}!}.
\end{equation}
Finally, we have the relation:
\begin{equation}
F^{abcde}_{00100 \vert 00v00} = c \,\delta^a_0 \delta^b_0 \delta^c_{v+1} \delta^d_0 \delta^e_0,
\end{equation}
from which we can lower the $v$-index as:
\begin{equation}
P_{00100}P_{00(v-1)00} = v P_{00v00}.
\end{equation}
This leads to the final expression for the dual basis:
\begin{equation}
\boxed{
P_{tuvwx} = \frac{P_{10000}^t}{[t]_{\tilde{q}^{-2}}!}\frac{P^u_{01000}}{[u]_{q^{-2}}!} \frac{P^v_{00100}}{v!}\frac{P^w_{00010}}{[w]_{q^{2}}!} \frac{P^x_{00001}}{[x]_{\tilde{q}^{2}}!}.
}
\end{equation}
Inserting now finally this expression and \eqref{eq:ini} into the object $g$, we find the promised representation of the quantum matrix group studied in the main text:
\begin{equation}
g = \sum P_{tuvwx}X^{tuvwx} = e_{\tilde{q}^{-2}}^{i \gamma^{1/b^2}\tilde{F}} \, e_{q^{-2}}^{i \gamma F} \,  e^{2\phi H} \, e_{q^{2}}^{-i \beta E} \, e_{\tilde{q}^2}^{-i \beta^{1/b^2}\tilde{E}}.
\end{equation}
The pairs of generators ($E,F,H$) and ($\tilde{E},\tilde{F},\tilde{H}\equiv b^2 H$) satisfy the quantum algebra relations
\begin{align}
\label{eq:algg}
[H,E]&=E,\quad [H,F]=-F,\quad [E,F] = \frac{q^{2H}-q^{-2H}}{q-q^{-1}}, \\
[\tilde{H},\tilde{E}]&=\tilde{E},\quad [\tilde{H},\tilde{F}]=-\tilde{F},\quad [\tilde{E},\tilde{F}] = \frac{\tilde{q}^{2\tilde{H}}-\tilde{q}^{-2\tilde{H}}}{\tilde{q}-\tilde{q}^{-1}}, \\
[E,\tilde{E}] &= 0 = [E,\tilde{F}], \quad [F,\tilde{E}] = 0 = [F,\tilde{F}],
\end{align}
which can be proven explicitly again using the co-product relations \eqref{eq:co1}-\eqref{eq:co2}.\footnote{The only relation here that might require some explanation is $[E,F]$ and its dual. To derive this relation, one needs to use the fact that the product $EF$ can be expanded into a sum of terms $P_{00v00}$ for all odd $v$, plus a term $P_{01010}$. This last term cancels the $-FE$ term. The sum of the first terms precisely give the series expansion of the RHS $\frac{q^{2H}-q^{-2H}}{q-q^{-1}}$ in $2H$, upon using the algebra relations $[\beta,\phi] = -\pi i b \beta$ and $[\gamma,\phi] = -\pi i b \gamma$. The dual relation $[\tilde{E},\tilde{F}]$ is analogous where extra factors of $b^{-2}$ are sprinkled throughout the computation.} The first and second line are the separate U$_q(\mathfrak{sl}(2,\mathbb{R}))$ and U$_{\tilde{q}}(\mathfrak{sl}(2,\mathbb{R}))$ relations, whereas the last line states that both of these quantum algebras commute.\footnote{The reader might worry that the $H$ and $\tilde{H}$ elements of each algebra do not commute with the dual algebras. This is why one usually defines the modular double using the exponentiated generators $K^2 = q^{2H}$ and $\tilde{K}^2 \equiv \tilde{q}^{2\tilde{H}}$ instead, as introduced in \eqref{PT}. In this case, the $K^2$ and $\tilde{K}^2$ generators do commute with their respective dual quantum algebras. See e.g. \cite{Kharchev:2001rs} for additional comments on this. Hence our construction and quantum algebra relations are consistent with these considerations, and provide the ``infinitesimal'' definition of the modular double.}

\section{Mellin transform and Gamma functions}
\label{app:A}
First the classical case. The Gamma function is defined by its Euler definition, and inverse:
\begin{equation}
\Gamma(z) = \int_0^{+\infty}dt \, t^{z-1}e^{-t}, \qquad
e^{z} = \frac{1}{2\pi i} \int_{i\mathbb{R}}ds \, \Gamma(s) (-z)^{-s}.
\end{equation}
We can prove this last formula directly by contour deforming the $s$-integral to the left half plane, and picking up all poles of the $\Gamma$-function with residue Res$_{s = -n} \Gamma(s) = (-)^n/n!$, so the RHS indeed becomes
\begin{equation}
\sum_{n=0}^{+\infty} \frac{z^{n}}{n!}  = e^{z}.
\end{equation}
The $q$-exponential introduced above has an analogous inverse Mellin transform in terms of a $q$-Gamma function:
\begin{equation}
e_q^{z} \equiv \frac{1}{2\pi i} \int_{i\mathbb{R}}ds \hat{\Gamma}_q(s) (-z)^{-s},
\end{equation}
which we define by this relation. The $q$-Gamma function is meromorphic on the complex plane with simple poles at $s=-n, \, n \in \mathbb{N}$, with residues Res$_{s = -n} \hat{\Gamma}_q(s) = (-)^n/[n]_q!$. Indeed, contour deforming the RHS to the left half-plane, we similarly get:
\begin{equation}
\sum_{n=0}^{+\infty}\frac{z^{n}}{[n]_q!} \equiv e_q^{z},
\end{equation}
which is the $q$-exponential function.
This $\hat{\Gamma}_q$ function satisfies its defining relation:
\begin{equation}
\hat{\Gamma}_q(z+1)=-[-z]_q \hat{\Gamma}_q(z).
\end{equation}
This holds for the ``compact'' $q$-deformation. We however are interested in the ``non-compact'' case. Instead of the $q$-Gamma function, the relevant quantity is then the double sine function $S_b(x)$.

\mciteSetMidEndSepPunct{}{\ifmciteBstWouldAddEndPunct.\else\fi}{\relax}
\bibliographystyle{utphys}
{\small \bibliography{references}{}}

\end{document}